\documentclass[10pt,a4paper,aps,prb,twocolumn,superscriptaddress,showpacs]{revtex4-1}

\usepackage{amsbsy}
\usepackage{amsfonts}
\usepackage{amsmath}
\usepackage{amssymb}
\usepackage{array}
\usepackage{bbding}
\usepackage{bbm}
\usepackage{bm}
\usepackage{epsfig}
\usepackage{fixmath}
\usepackage{float}
\usepackage[T1]{fontenc}
\usepackage{graphicx,rotating}
\usepackage{multirow}
\usepackage{placeins}
\usepackage[svgnames]{xcolor}
\usepackage{tikz}
\usepackage{lipsum}
\usepackage{curves}
\usepackage{times}

\DeclareMathAlphabet{\mathantt}{OT1}{antt}{li}{it}
\DeclareMathAlphabet{\mathpzc}{OT1}{pzc}{m}{it}

\providecommand*{\I}{\mathrm{i}}                           %% imaginary unit i
\providecommand*{\bra}[1]{\langle#1|}                      %% bra vector
\providecommand*{\ket}[1]{|#1\rangle}                      %% ket vector
\newcommand{\tr}{\mathrm{tr}}													%% Trace
\renewcommand{\vec}[1]{\mathbold{#1}}
\newcommand{\nn}{\nonumber}
\renewcommand{\d}{\mathrm{d}}

\newcommand{\bok}[3]{\left<\right.\hspace{-0.5ex}{#1}\left.\hspace{-0.5ex}\right|{#2}\left|\right.\hspace{-0.5ex}{#3}\left.\hspace{-0.5ex}\right>}

\newcommand{\mw}[1]{\left<\right.\hspace{-0.5ex}{#1}\left.\hspace{-0.5ex}\right>}

\definecolor{ao(english)}{rgb}{0.0, 0.5, 0.0}

\newcommand{\vF}{v_{\mathrm{F}}}

\newcommand{\Int}[1][-5pt]{\int\limits_{\begin{picture}(16,3)(-8,-3)%
		\put(0,0){\curve(-3,0,-8,0)\curve(3,0,8,0)}%
		\put(8,0){\curve(0,0,-1.5,1.5)\curve(0,0,-1.5,-1.5)}%
		\put(0,0){\arc(-3,0){180}}\put(0,0){\makebox(0,0){$\cdot$}}%
		\end{picture}}\hspace*{#1}}
		
\newcommand{\Exp}[1]{\mathrm{e}^{\mbox{\footnotesize$#1$}}}

\begin{document}

%\preprint{}

\title{First-principles quantum corrections for carrier correlations\\ in double-layer two-dimensional heterostructures}


\author{Martin-Isbj\"orn~Trappe}
\email[]{martin.trappe@quantumlah.org}
%\thanks{}
\affiliation{Centre for Advanced 2D Materials and Graphene Research Centre, National University of Singapore, 6 Science Drive 2, Singapore 117546, Singapore}
\affiliation{Centre for Quantum Technologies, National University of Singapore, 3 Science Drive 2, Singapore 117543, Singapore}

\author{Derek~Y.~H.~Ho}
\email[]{derek_ho@nus.edu.sg}
%\thanks{}
\affiliation{Centre for Advanced 2D Materials and Graphene Research Centre, National University of Singapore, 6 Science Drive 2, Singapore 117546, Singapore}

\author{Shaffique Adam}
\email[]{shaffique.adam@yale-nus.edu.sg}
%\thanks{}
\affiliation{Centre for Advanced 2D Materials and Graphene Research Centre, National University of Singapore, 6 Science Drive 2, Singapore 117546, Singapore}
\affiliation{Yale-NUS College, 16 College Avenue West, Singapore 138527, Singapore}

%\author{\large \footnotemark[1]\hspace{1ex}, \large \footnotemark[1]\hspace{1ex}, \large \footnotemark[2], and \large \footnotemark[1]\hspace{1ex}\\
% \scriptsize\footnotemark[1]\hspace{1ex}Centre for Quantum Technologies, National University of Singapore\\[-0.7em]
% \scriptsize\footnotemark[2]\hspace{1ex}Department of Physics, University of Konstanz, Germany}


\date{\today}

\begin{abstract}
We present systematic ab initio calculations of the charge carrier correlations between adjacent layers of two-dimensional materials in the presence of both charged impurity and strain disorder potentials using the examples of monolayer and bilayer graphene. For the first time, our analysis yields unambiguous first-principles quantum corrections to the Thomas--Fermi densities for interacting two-dimensional systems described by orbital-free density functional theory. Specifically, using density-potential functional theory, we find that quantum corrections to the quasi-classical Thomas-Fermi approximation have to be taken into account even for heterostructures of mesoscopic size. In order for the disorder-induced puddles of electrons and holes to be anti-correlated at zero average carrier density for both layers, the strength of the strain potential has to exceed that of the impurity potential by at least a factor of ten, with this number increasing for smaller impurity densities. Furthermore, our results show that quantum corrections have a larger impact on puddle correlations than exchange does, and they are necessary for properly predicting the experimentally observed Gaussian energy distribution at charge neutrality.\\

%\\[1em] \blue{Things to do in blue.} \green{New parts in green.} \red{Corrections in red.}\\

%PhySH: Two-dimensional electron systems, Density functional approximations, Semiclassical physics, Graphene, Heterostructures, Coulomb drag, Semiconducters

PhySH: Two-dimensional electron system, Density functional approximations, Heterostructures

\end{abstract}

\maketitle

\section{\label{Intro}Introduction}

The simulation of two-dimensional (2D) materials and prediction of their properties has become a mainstay of materials science over the past decade, with the promise and realization of valuable applications in both industrial technology and fundamental research~\cite{geim_van_2013}. Theoretical and computational methods for 2D materials have been advanced into a sophisticated machinery that enables researchers to deal with ever more realistic settings~\cite{rodriguez-vega_ground_2014}. The widely used Kohn-Sham density functional theory (KS-DFT)~\cite{KohnSham1965,Dreizler1990} presents one particularly popular ab initio approach with the capability of accurately handling hundreds of interacting particles (and up to thousands of atoms in cases where linear-scaling methods apply \cite{Goedecker1999,Soler2002,Bowler2012,Fox2014,Aarons2016,Cole2016}). The development of functionals for 2D systems has been lagging behind that of their 3D counterparts for various reasons: Some of the most heavily used 3D KS-DFT functionals are not bounded from below \cite{Chiodo2012,Kaplan2018} when the 2D limit is approached. This stems from the improper scaling behaviour of the density \cite{Chiodo2012}. Furthermore, even consistent first-order gradient corrections of the kinetic energy density (used in developing meta-generalized gradient approximations for KS-DFT) were unknown for 2D fermion systems until recently \cite{Holas1991,Brack2003,Trappe2016}. Nonetheless, considerable progress has been made and alternative derivations of some of those 2D functionals have been obtained since the turn of the millenium \cite{Pollack2000,Reimann2002,Pittalis2007,Pittalis2008,Constantin2008,Pittalis2009,Pittalis2009b,Pittalis2010,Raesaenen2010b,Vilhena2014}.

However, a systematic ab inito methodology that is universally applicable and scales favorably with particle number, thereby enabling high-throughput computations of mesoscopic systems, is not yet available --- with orbital-free density functional theory (OF-DFT) being the suspected saviour for almost a century~\cite{Thomas1927,Fermi1927,Gross2011,Burke2012,Karasiev2013,GrossBurke2015,Witt2018}. While KS-DFT scales cubically with particle number in generic settings, OF-DFT scales linearly, and sub-linear scaling can be achieved in special cases. Functional development, in particular concerning the kinetic energy functional \cite{Wang1992,DellaSala2015,Xia2015}, and implementations of OF-DFT have gained momentum in recent years \cite{Karasiev2012b,Karasiev2013,Chen2015,Das2015,Chen2016,Mi2016,Constantin2018,Witt2018}, also in conjunction with other techniques like ab initio molecular dymanics \cite{Gonzalez2008}. Chemical accuracy is approached in selected cases \cite{Xia2012,Borgoo2014,EspinosaLeal2015}. If quantum effects play a minor role or if the considered system is largely homogeneous, OF-DFT can also be used in its most basic form, the Thomas--Fermi (TF) approximation. For instance, the effect of exchange on large disordered systems with long-range interactions was studied in Ref.~[\onlinecite{Rossi2008}] using OF-DFT in TF approximation. The TF model is not only of historical significance, but presents, as an exact constraint for homogeneous systems and in the limit of infinite nuclear charges \cite{Lieb1977}, an important base line for benchmarking proposed systematic density functional improvements. To what extent then do corrections to the TF approach play a crucial role or dominate over exchange effects in 2D materials? (They do, indeed, for a number of relevant fermionic systems, ranging from atomic Fermi gases to molecules and single atoms.)

The most severe obstacle for OF-DFT in taking over as the workhorse of theoretical chemistry and materials science is the lack of accurate, reliable, systematic, and preferably universal quantum corrections to the quasi-classical TF approximation, in particular for the kinetic energy of low-dimensional systems~\cite{Holas1991,Brack2003,Salasnich2007,Trappe2016,Trappe2017}. While ad-hoc corrections to the quasi-classical limit and heuristic approximations are available for kinetic energy and particle density of low-dimensional systems~\cite{vanZyl2013,vanZyl2014}, successful derivations of systematic and consistent corrections are scarce~\cite{Burke2015,TrDeMu2015,Trappe2016,Trappe2017,Chau2018}. One promising route towards systematic orbital-free quantum corrections is provided by density-potential functional theory (DPFT)~\cite{Berge1988,Berge1992,CinalBerge1993,Trappe2016,Trappe2017,Chau2018}, a more flexible reformulation of the original Hohenberg-Kohn DFT~\cite{HohenbergKohn1964,Dreizler1990}, which circumvents the need for an explicit kinetic-energy density-functional and provides natural ways for systematic semiclassical expansions.

In this article we explore the applicability of DPFT for 2D materials by assessing quantum-corrections to the TF approximation for double-layer heterostructures of mono- and bilayer graphene. Of particular interest to us are situations that are not easily tackled with orbital-based techniques, for example 2D material sheets of mesoscopic size that are subjected to aperiodic disorder potentials. Such situations are for example of current interest in studies on Coulomb drag~\cite{narozhny_coulomb_2016} where there exists an unsettled controversy as to whether the behavior of drag measured in experiment~\cite{gorbachev_strong_2012} is due to correlation~\cite{song_energy-driven_2012} or anti-correlation~\cite{ho_theory_2018} between the density fluctuations of the layers.

Our work contributes in several ways to answering some of the questions raised above. Sections \ref{DPFT} and \ref{SCsim} provide the computational framework for obtaining quantum-corrected carrier-densities of 2D materials using DPFT. The expressions for the semiclassical particle densities developed here enable us to decide whether or not the quasi-classical TF approximation is sufficient for describing at least conglomerate properties like average inter-layer correlations of heterostructures. Section~\ref{Bilayersetup} introduces the generic double-layer system, with both layers subjected to one layer of charged impurities, while only one of the layers is strained. Charge and strain disorder potentials are expected to compete in creating correlated (from charged impurities) and anticorrelated (from strain) carrier densities in the two layers. Our model setup is designed to extract the strain strengths required for switching between correlation and anticorrelation. In Sec.~\ref{Results} we apply our new approach to double-monolayer graphene and double-bilayer graphene. We discuss whether or not the electron-hole puddles of both layers, interacting electrostatically, require a self-consistent inter-layer treatment. Finally, we analyse the effects of quantum-corrections and exchange energy on the correlations with the aid of phase diagrams that chart the correlation measures as functions of impurity density, carrier density, and ratios of strain and charge disorder. The appendix gathers background information on the units, system parameters, correlation measures, and numerical procedures employed here.

\section{\label{DPFT}Density-potential functional theory}
Instead of resorting to the computationally demanding orbital-based Kohn-Sham DFT, we make use of orbital-free density-potential functional theory (DPFT)~\cite{Berge1988,Berge1992}. It is formally equivalent to the Hohenberg-Kohn formulation, but makes systematic improvements upon the TF approximation technically feasible --- in particular for low-dimensional systems.

Specifically, by Legendre-transforming $E_{\mathrm{kin}}[n]$, the kinetic energy functional of the particle density $n(\vec r)$, w.r.t.~the new variable ${V(\vec r)=\mu-\frac{\delta E_{\mathrm{kin}}[n]}{\delta n(\vec r)}}$, we recast the total energy of an interacting quantum system with interaction energy $E_{\mathrm{int}}[n]$,
\begin{align}\label{Energynmu}
E[n,\mu]&=E_{\mathrm{kin}}[n]+E_{\mathrm{ext}}[n]+E_{\mathrm{int}}[n]\nn\\
&\quad+\mu\left(N-\int(\d\vec r)\,n(\vec r)\right),
\end{align}
as the density-potential functional
\begin{align}\label{EnergyVnmu}
E[V,n,\mu]&=E_1[V-\mu]-\int(\d\vec r)\,n(\vec r)\,\big(V(\vec r)-V_{\mathrm{ext}}(\vec r)\big)\nn\\
&\quad+E_{\mathrm{int}}[n]+\mu N.
\end{align}
Here, the external potential $V_{\mathrm{ext}}(\vec r)$ yields the external energy $E_{\mathrm{ext}}[n]$, and the particle number $N$ is enforced via the Lagrange multiplier $\mu$, viz.~the chemical potential. From Eq.~(\ref{EnergyVnmu}) we obtain the ground-state solutions of the three variables $V$, $n$, and $\mu$ by self-consistently solving
\begin{align}
n(\vec r)&=\frac{\delta E_1[V-\mu]}{\delta V(\vec r)},\label{nasdef}\\
V(\vec r)&=V_{\mathrm{ext}}(\vec r)+\frac{\delta E_{\mathrm{int}}[n]}{\delta n(\vec r)}, \label{Vdef}\\
N&=\int(\d\vec r) n(\vec r) \label{Nmu}.
\end{align}
Equation~(\ref{Nmu}) is obtained by combining $\partial E[V,n,\mu]/\partial\mu$ with Eq.~(\ref{nasdef}) and reveals the particle number constraint in Eq.~(\ref{Energynmu}).

Equations~(\ref{nasdef})--(\ref{Nmu}) are exact and reminiscent of the KS scheme, but without the need of orbitals. However, the noninteracting case aside, we have to approximate the unknown potential functional ${E_1[V-\mu]}$; here the subscript indicates that $E_1$ can be written in terms of a single-particle trace over a function of the single-particle Hamilton operator \cite{Berge1992}. We also have to provide the equally important interaction energy $E_{\mathrm{int}}[n]$ as an explicit functional of the particle density $n$. Approximate particle densities follow directly from approximations of ${E_1[V-\mu]}$ (or, rather, its functional derivative) for any given potential $V$. As is evident from Eq.~(\ref{Vdef}), $V$ constitutes an effective single-particle potential with interaction effects effectively included for any given density $n$

Following Refs.~[\onlinecite{BGE1984,Berge1992,CinalBerge1993,Trappe2016,Trappe2017,Chau2018}] we approximate $E_1$ by its noninteracting version as the single-particle trace
\begin{align}\label{E1}
E_1[V-\mu]=\tr\{(H_1-\mu)\,\eta(\mu-H_1)\},
\end{align}
where ${H_1=H_1(\vec R,\vec P)=T(\vec P)+V(\vec R)}$ is a single-particle Hamiltonian with dispersion relation $T$ and potential energy $V$, while the trace includes the degeneracy factor $g$. For example, ${g=4}$ accounts for the spin and valley multiplicity of unpolarized charge carriers in the cases of mono- and bilayer graphene. $\vec R$ and $\vec P$ are the position and momentum operators, respectively, and $\eta(\,)$ denotes the step function.

The explicit expression for $E_1$ in Eq.~(\ref{E1}) results in an explicit expression for the particle density in terms of arbitrary functions $V(\vec r)$ via Eq.~(\ref{nasdef}). The approximate nature of Eq.~(\ref{E1}) aside, the exact particle density including all quantum corrections is thereby obtained for any specified interaction energy $E_{\mathrm{int}}[n]$ and without reference to orbitals. Specifically, Eqs.~(\ref{nasdef}) and (\ref{E1}), together with the Fourier transform of the step function, yield the particle density\footnote{The contour integration circumvents the singularity at ${t=0}$ in the lower half-plane.}
\begin{align}\label{nPropagator}
%n(\vec r)=g\int_{-\infty}^\mu\d E\int\frac{\d t}{2\pi\hbar}\,\mathrm{e}^{\frac{\I}{\hbar}E\,t}\bok{\vec r}{\mathrm{e}^{-\frac{\I}{\hbar}H\,t}}{\vec r},
n(\vec{r})=g\Int\frac{\d t}{2\pi\I t}\,\Exp{\frac{\I t}{\hbar}\mu}\bra{\vec{r}}\Exp{-\frac{\I t}{\hbar}H_1}\ket{\vec{r}},
\end{align}
see Refs.~[\onlinecite{Golden1957b,Golden1960,Light1973,Lee1975,Chau2018}].

%\begin{widetext}
\begin{table*}[t]
\centering
\begin{tabular}{c|c|c}
%\begin{tabular}{|c|c|c|}
%\hline
$T(\vec p)$ & $n_{\mathrm{TF}}(\vec r)$ & $n_3(\vec r)$ \\
& & \\[-2.0ex]
\hline\hline\\[-2.0ex]
$\vF|\vec p|$ & $\frac{g}{4\pi}K(\vec r)^2$ & $\frac{g}{4\pi^2}\int(\d\vec s)\frac{K(\vec r+\vec s)^3}{|\vec s|}\,\mathrm{J}_1\big(2|\vec s|\,K(\vec r+\vec s)\big)$ \\
& & \\[-2.0ex]
\hline\\[-2.0ex]
$\frac{\vec p^2}{2m}$ & $\frac{g}{4\pi}\sigma(\vec r)$ & $\frac{g}{4\pi^2}\int(\d\vec s)\frac{\sigma(\vec r+\vec s)}{|\vec s|^2}\,\mathrm{J}_2\big(2|\vec s|\sqrt{\sigma(\vec r+\vec s)}\,\big)$ \\
%\hline
\end{tabular}
\caption{\label{TableDensities}Thomas--Fermi density $n_{\mathrm{TF}}(\vec r)$ and quantum-corrected density $n_3(\vec r)$ for linear and quadratic dispersion in 2D from evaluating Eq.~(\ref{nPropagator}), with degeneracy factor $g$, Bessel functions $\mathrm{J}_\alpha(\,)$, ${\nu(\vec z)=[\mu-V(\vec z)]_+}$, ${K(\vec z)=\nu(\vec z)/(\hbar\vF)}$, ${\sigma(\vec z)=2m\,\nu(\vec z)/\hbar^2}$, and $[x]_+$ denoting $x\,\eta(x)$. We recover the TF densities from their quantum-corrected successors upon replacing ${K(\vec r+\vec s)}$ and ${\sigma(\vec r+\vec s)}$ by their local versions $K(\vec r)$ and $\sigma(\vec r)$, respectively. The case of quadratic dispersion is dealt with in Ref.~[\onlinecite{Chau2018}]; see Appendix~\ref{Appendixn3} for the derivations in the case of linear dispersion.}
\end{table*}
%\end{widetext}

We seek to approximate the time evolution operator ${U=\mathrm{e}^{-\frac{\I t}{\hbar}H_1}}$ systematically via split-operator methods, for example of the Suzuki-Trotter type~\cite{Hatano2005,Chau2018}. The quasi-classical approximation ${U\approx U_2=\mathrm{e}^{-\frac{\I t}{\hbar}T(\vec P)} \mathrm{e}^{-\frac{\I t}{\hbar}V(\vec R)}}$ recovers the quasi-classical TF density $n_{\mathrm{TF}}$, while ${U_3=\mathrm{e}^{-\frac{\I t}{2\hbar}T(\vec P)} \mathrm{e}^{-\frac{\I t}{\hbar}V(\vec R)}\mathrm{e}^{-\frac{\I t}{2\hbar}T(\vec P)}}$ produces the first quantum-corrected density $n_3$ in a series of expressions that utilize higher-order factorizations~\cite{Chau2018} of $U$. We give the corresponding 2D densities for linear and quadratic dispersion in Table~\ref{TableDensities} and outline the derivation of $n_3$ for the case of linear dispersion in Appendix~\ref{Appendixn3}. In contrast to $n_{\mathrm{TF}}(\vec r)$, which is restricted to classically allowed regions of the potential and only depends on the local value $V(\vec r)$, $n_3(\vec r)$ samples $V$ in an extended region and exhibits evanescent tails beyond the quantum-classical border.

\section{\label{SCsim}Self-consistent simulation of disordered 2D materials}

We target 2D systems with chemical potential in the vicinity of the Dirac point (the point where valence and conduction bands touch) for graphene (effective bilayer graphene). The usual tight binding approach absorbs the lattice structure in an effective Hamiltonian and yields noninteracting quasiparticles in a homogeneous (${V_{\mathrm{ext}}(\vec r)=0}$) environment. The electronic structures, viz.~atoms, of the materials are thus not modeled explicitly here. The single-particle energies, viz.~band structure, associated with such quasiparticles is the dispersion relation $T(\vec p)$ whose operator version appears in Eq.~(\ref{E1}). $T(\vec p)$ can be an arbitrary function, but for the purpose of this work we shall restrict ourselves to the analytically more tractable cases of linear and quadratic dispersion\footnote{Hence, the scope of our investigation also extends to energies further away from the Dirac point (band gap) as long as $T(\vec p)$ can be considered linear (quadratic).}.

In the following we outline the procedures involved for arriving at the ground state solutions of Eqs.~(\ref{nasdef})--(\ref{Nmu}); further details are provided in Appendix~\ref{AppendixSCloop}. Upon adding external potentials $V_\mathrm{C}(\vec r)$ and $V_\mathrm{S}(\vec r)$ that model charged impurites and strain, respectively, we initiate the self-consistent loop of Eqs.~(\ref{nasdef})--(\ref{Nmu}) by evaluating the density (denoted $n_-(\vec r)$ for the quasiparticles that follow the dispersion of the conduction band) with the external potential
\begin{align}\label{V-}
V_{\mathrm{ext}}(\vec r)=V_\mathrm{C}(\vec r)+V_\mathrm{S}(\vec r)
\end{align}
for these conduction quasiparticles, see Appendix~\ref{AppendixDisorder} for details. Since no interactions are included at this stage, the effective potential is ${V_-(\vec r)=V_{\mathrm{ext}}(\vec r)}$. The density $n_+(\vec r)$ of valence quasiparticles, which follow the inverted dispersion, e.g. $T(\vec p)=-\vF|\vec p|$ in the case of graphene, is built from the same density expression as $n_-(\vec r)$ but takes as an input the inverted potential $2\,(\mu+\Delta)-V_{\mathrm{ext}}(\vec r)$, with an optional bandgap $\Delta$ (in case of mono- and bilayer graphene, we have ${\Delta=0}$). The effective potential for the valence quasiparticles reads
\begin{align}\label{V+}
V_+(\vec r)=2\,(\mu+\Delta)-V_-(\vec r)
\end{align}
and equals the external potential for the valence quasiparticles if interactions are omitted. The such obtained carrier density ${n(\vec r)=n_-(\vec r)-n_+(\vec r)}$ updates the effective potential via the interaction contribution in Eq.~(\ref{Vdef}).

As an approximate interaction energy $E_{\mathrm{int}}[n]$ for the quasiparticles, we employ the regularized Hartree term for the Coulomb energy,
\begin{align}\label{HartreeEnergy}
E_{\mathrm{H}}[n]=\frac{W}{2}\int(\d\vec r)(\d\vec r')\frac{n(\vec r)\,n(\vec r')}{\mathrm{max}(|\vec r-\vec r'|,b)},
\end{align}
where $b$ is half the lattice constant of the numerical implementation and $W=\hbar\vF r_\mathrm{s}$, with the ratio $r_\mathrm{s}$ of Coulomb potential energy and kinetic energy for graphene. Upon functional differentiation, Eq.~(\ref{HartreeEnergy}) leads to the Hartree potential\footnote{Equation (\ref{HartreePotential}) can be regarded as a simplified version of the smooth models for the Coulomb potential discussed in Ref.~[\onlinecite{GonzalezEspinoza2016}].}
\begin{align}\label{HartreePotential}
V_{\mathrm{H}}(\vec r)=W\int(\d\vec r)\frac{n(\vec r')}{\mathrm{max}(|\vec r-\vec r'|,b)}
\end{align}
as an approximate interaction contribution in Eq.~(\ref{Vdef}). Eyeing means of comparison and higher accuracy, we may supplement $E_{\mathrm{H}}$ with an exchange energy~\cite{Dirac1930}, leading to the exchange potential $V_{\mathrm{X}}$; see Appendix~\ref{AppendixExchange} for details.

\begin{figure}[htb!]
\begin{center}
\includegraphics[width=0.95\linewidth]{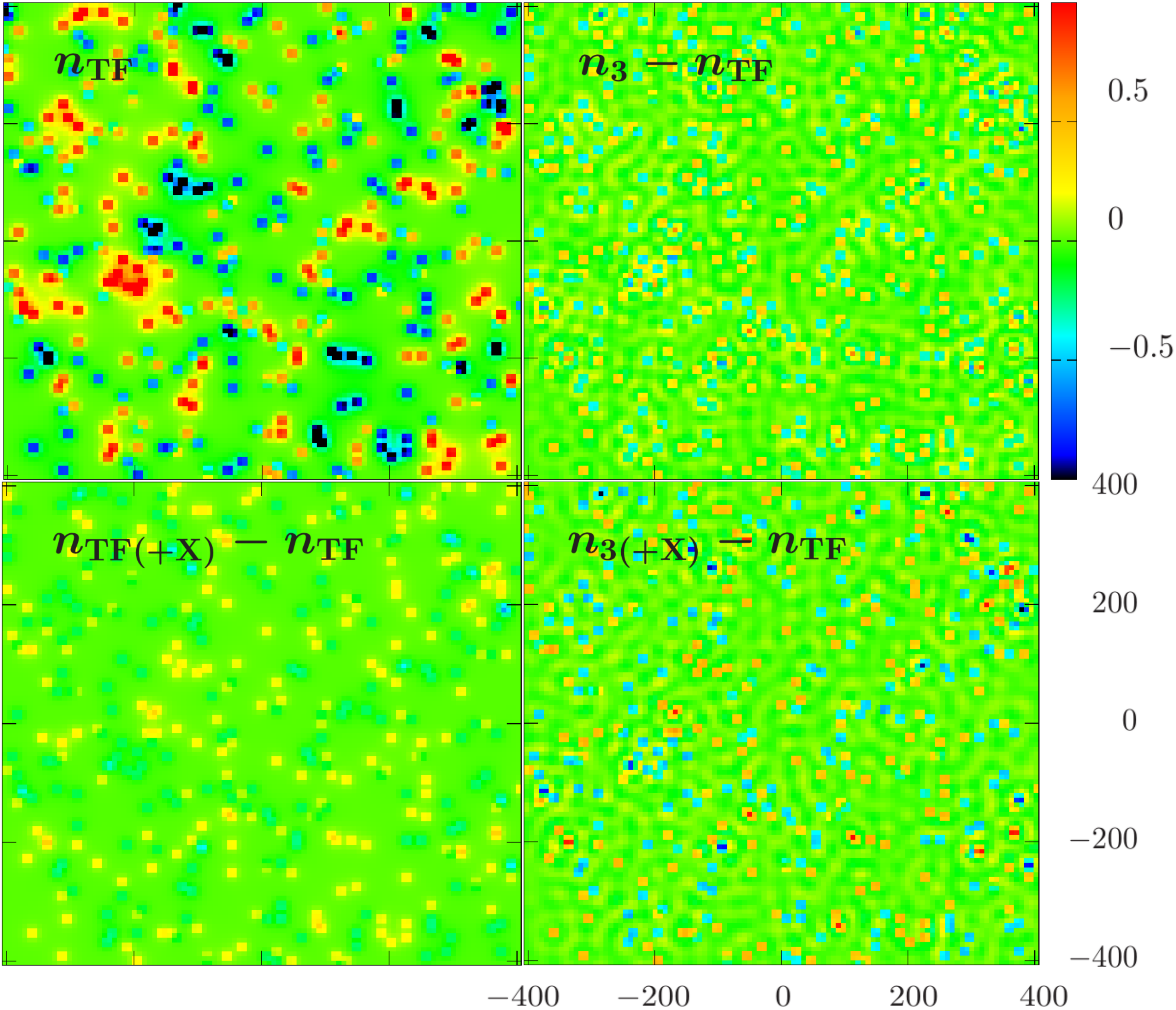}
\caption{\label{TFvsST3+X_monolayer} The spatial distributions of quasiparticle densities $n_{\mathrm{TF}}$ and $n_3$ of monolayer graphene visibly depend on exchange and quantum corrections. The upper left panel shows the Thomas--Fermi density $n_{\mathrm{TF}}$. The other three panels display the differences between $n_{\mathrm{TF}}$ and the quantum-corrected density $n_3$ (upper right panel), the TF density with the exchange potential of Eq.~(\ref{ExchangePotentialRossi}) included (lower left panel), and $n_3$ including exchange (lower right panel). The color bar refers to densities measured in units of ${0.001/l^2}$. The horizontal and vertical axes indicate spatial position, in the units of $l$ as labelled explicitly in the lower right panel. The same disorder realization, with average impurity density ${\bar{n}_{\mathrm{imp}} = 10^{12}/\mathrm{cm}^{2}}$, is used in all four panels. }
\end{center}
\end{figure}

The updated effective potential $V$ then determines a new quasiparticle density $n_-(\vec r)$ via Eq.~(\ref{nasdef}), thereby closing the self-consistent loop. This process is repeated until a predefined relative precision is reached (we find $10^{-6}$ to be sufficient) when comparing the local densities of subsequent loop iterations. The chemical potential is adjusted in each iteration to enforce a given particle number, viz.~average carrier density. Figure~\ref{TFvsST3+X_monolayer} highlights the differences in the converged quasiparticle densities $n_{\mathrm{TF}}$ and $n_3$ of a single graphene layer with and without exchange. Both exchange and quantum corrections tend to decrease the peaks of the density landscape. This effect is well-known in the case of exchange~\cite{Rossi2008}, and a smoothening of the carrier density in an external potential $V_\mathrm{C}$ is to be expected when tunneling starts to play a role with the inclusion of quantum corrections. In fact, when comparing the corresponding densities in Fig.~\ref{TFvsST3+X_monolayer} we find that quantum corrections considered in this work can dominate over exchange effects.

\begin{widetext}

\begin{figure}[htb!]
\begin{center}
\includegraphics[width=0.405\linewidth]{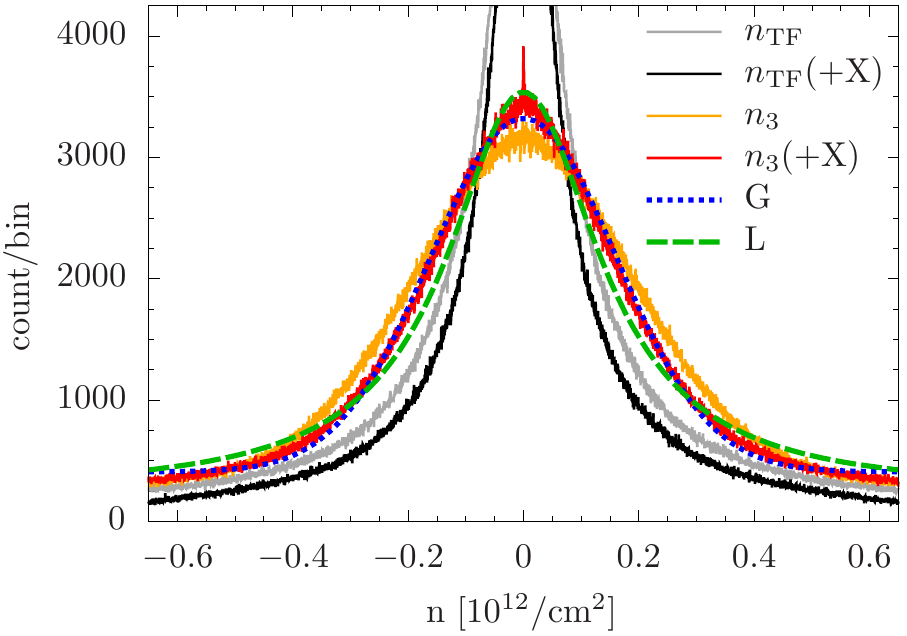}
\includegraphics[width=0.4\linewidth]{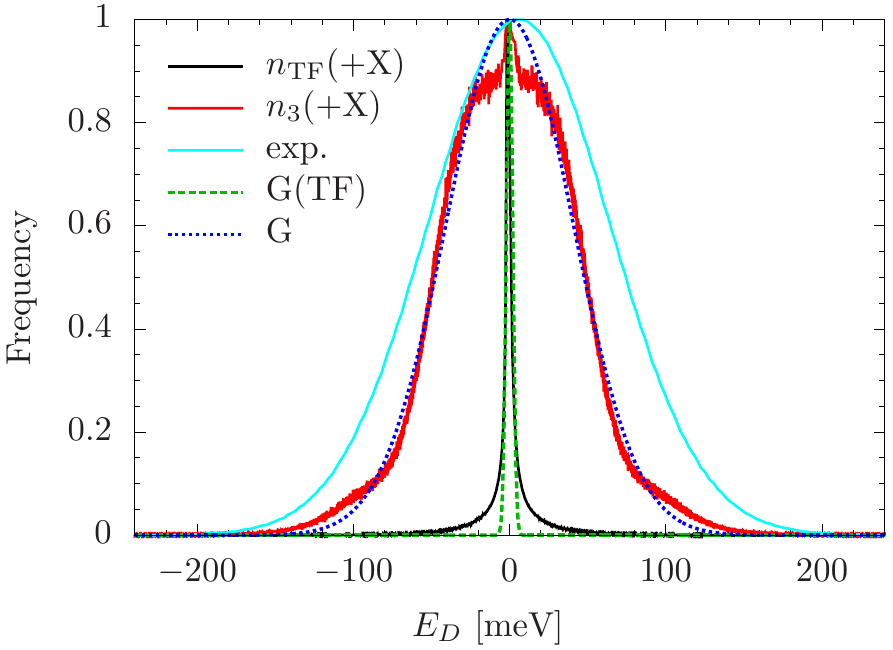}
\caption{\label{DisorderAverages_Histogram}Quantum corrections are crucial for obtaining the experimentally observed Gaussian energy distribution function for monolayer graphene on SiO$_2$ at charge neutrality. Left: Density histograms for $n_{\mathrm{TF}}$ and $n_3$, with and without exchange (X), for 500 disorder realizations with ${\mw{n_j}=0}$. We choose the bin widths for the densities such that 100 counts per bin are obtained on average. In contrast to the TF approximation, the quantum-corrected density distribution $n_3$ (including exchange) is captured reasonably well by a Lorentzian and even better by a Gaussian fit. Right: Translating local densities into energies ${E_D=\mathrm{sgn}\big(n(\vec r)\big)\,\hbar v_{\mathrm{F}}\,\sqrt{\pi|n(\vec r)|}}$, with the signum function $\mathrm{sgn}(\,)$, we find that the local quantum-corrected energies $E_D$ follow a Gaussian distribution that resembles the Gaussian fit to the experimental data `exp.' (extracted from Ref.~[\onlinecite{xue_scanning_2011}]) remarkably well. This is in stark contrast to the results of the TF approximation. We convert counts/bin into frequencies by renormalizing the histograms with the maxima of the Gaussian fits `G' and `G(TF)', respectively.}
\end{center}
\end{figure}

\end{widetext}

The analysis of disorder averages reveals a striking instance of this observation. As observed in Fig.~\ref{TFvsST3+X_monolayer} for a single disorder realization, $n_3$ and $n_{\mathrm{TF}}$ differ in their density distribution function. For $n_3$ the integration over a finite region in the disorder landscape, see Table~\ref{TableDensities}, tends to result in smoother densities compared with $n_{\mathrm{TF}}$. This effect can be quantified by density histograms compiled from many disorder realizations. The density histograms in Fig.~\ref{DisorderAverages_Histogram}, calculated for a graphene monolayer on SiO$_2$ with ${\bar{n}_{\mathrm{imp}} = 10^{12}/\mathrm{cm}^{2}}$, corroborate the snapshot of one disorder realization in Fig.~\ref{TFvsST3+X_monolayer}. Figures~\ref{TFvsST3+X_monolayer} and \ref{DisorderAverages_Histogram} highlight the (a priori) importance of not only addressing exchange but also quantum corrections beyond the TF approximation for quantitatively viable investigations of 2D materials via orbital-free DFT: Both the inclusion of exchange and of quantum corrections indicate less pronounced peaks of the carrier-density landscape. Compared to the TF approximation, the quantum corrections exhibit an averaging effect with broader density distributions that have relatively more weight on intermediate values of the density rather than a strong distribution maximum at zero density; cf. Fig.~\ref{DisorderAverages_Histogram} (left). Although exchange also shows visibly less pronounced densities in Fig.~\ref{TFvsST3+X_monolayer}, this effect stems from a global reduction in density variance rather than a redistribution of densities from very small towards intermediate values. The distributions of the quantum-corrected densities (including exchange) are captured by their Gaussian (`G') or Lorentzian (`L') fits (with offset) more accurately than the distributions resulting from the TF approximation. This observation is in line with experimental results for graphene that point towards Gaussian distributions of density and energy in the presence of charged disorder~\cite{martin_observation_2008,xue_scanning_2011}. In Fig.~\ref{DisorderAverages_Histogram} we compare our calculations `$n_{\mathrm{TF}}$(+X)' and `$n_3$(+X)', with their Gaussian fits `G(TF)' and `G', directly with the experimental data `exp.' from scanning tunnelling spectroscopy; cf.~Ref.~[\onlinecite{xue_scanning_2011}]. We find our quantum-corrected approach to predict the experimental data much better than what can be obtained from the TF approximation --- in both qualitative and quantitative terms. In view of this stark improvement over the TF approximation, we want to stress again that our quantum-corrected density expressions are based on first principles without adjustable parameters or fits, and rely solely on controlled approximations to quantum mechanics.

\section{\label{Bilayersetup}Double layer setup}

\begin{figure}[ht]
\begin{center}
\includegraphics[width=0.9\linewidth]{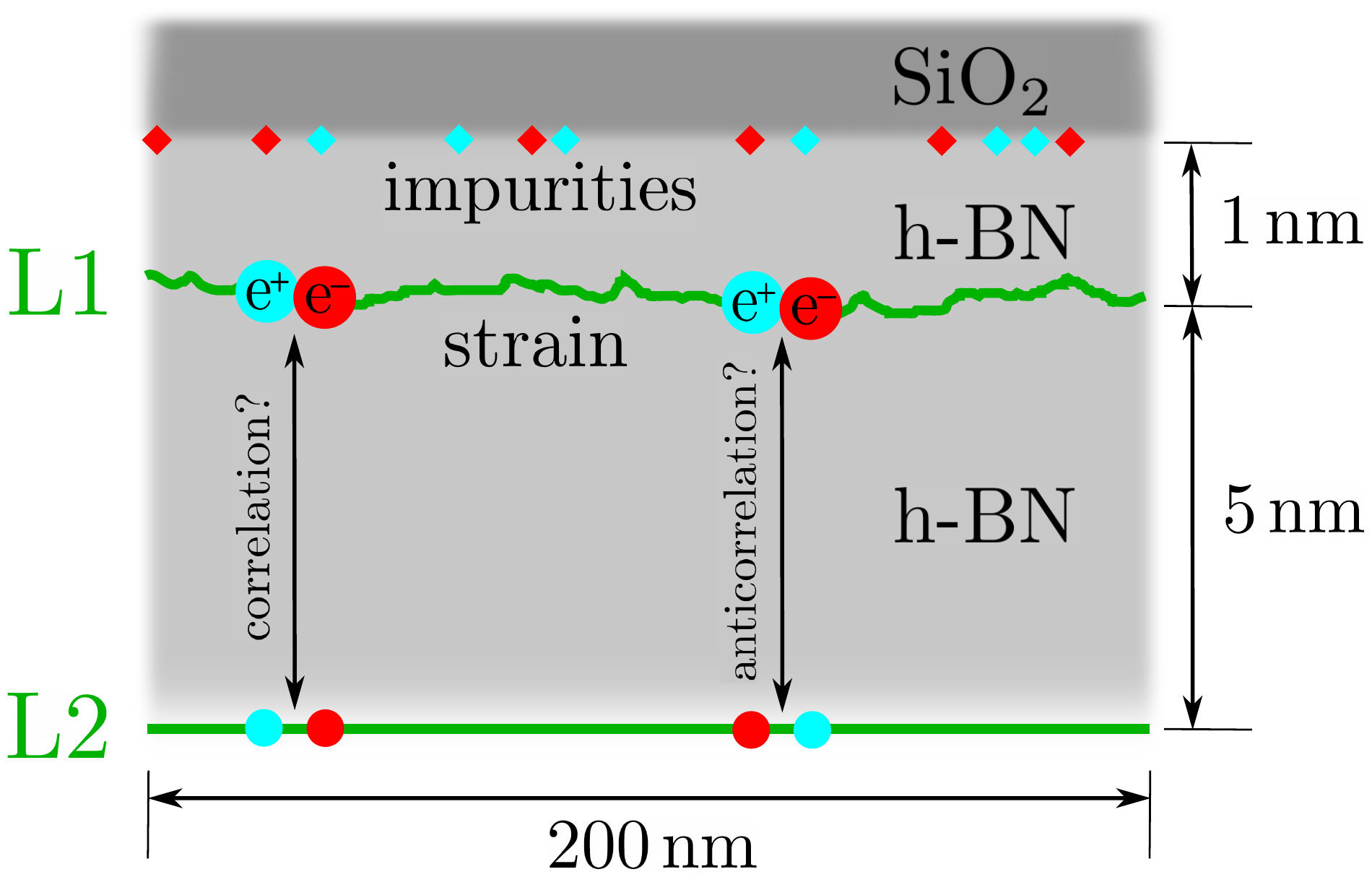}
\caption{\label{DoubleLayerSchematics}Schematics of the double-layer setup investigated here. Two finite-size monolayers are embedded in h-BN at a layer separation of $5\,$nm, while the SiO$_2$ substrate provides the charged impurities at an effective distance\cite{Tan2007,Chen2008b,Samaddar2016} of $1\,$nm from the first layer L1. For the purpose of revealing the impact of strain and charge disorder on the electron-hole-puddle correlations between L1 and L2, we model L2 as an unstrained clean layer. We use the same setup for analyzing both monolayer and bilayer graphene heterostructures.}
\end{center}
\end{figure}

The treatment of monolayers in the previous section forms our basis for the description of more complicated heterostructures. Figure~\ref{DoubleLayerSchematics} illustrates a two-layer system, where both layers L1 and L2 are sandwiched between h-BN and subjected to a charged impurity layer from the SiO$_2$ substrate. We expose L2 to the same charge disorder that affects L1, though at a larger separation, but refrain from adding disorder on L2 in order to avoid confusing inter-layer correlation effects with effects from independent disorder on L2. For the same reason we model the strain of L1 and the charge disorder by the identical type of disorder, albeit in different realizations. The layer separation of $5\,$nm suffices to justify a merely classical electrostatic interaction between L1 and L2, i.e., inter-layer tunneling of charge carriers can be neglected --- in contrast to intra-layer tunneling through the disorder potential landscape. The latter is missed by densities in TF approximation but captured (in part) via the higher-order Suzuki-Trotter factorizations.

In what follows we address the puddle correlations between L1 and L2 as a function of the ratio $R$ between the disorder strength of the strain and that of the charged impurities. Further details are provided in Appendix~\ref{AppendixDisorder}. The sum of $V_\mathrm{C}$ and $V_\mathrm{S}$ results in electron-hole puddles within the first layer L1, whose electrostatic potential adds to the external potential $V_\mathrm{C}$ for the charge carriers in the second layer\footnote{Strictly speaking, the electrostatic effects of L1 and L2 on each other have to be treated self-consistently, but the layer separation of 5nm ensures that the backreaction of L2 on L1 is of the order of $10^{-6}$ of the external potential of L1 and therefore negligible at the level of precision we are aiming for in this work.}. We expect maximal puddle correlation if ${R=0}$, that is, when no strain can obscure the then dominating effect of the charged impurities on both layers: Owing to the intra-layer Coulomb interaction, the puddles in L1 are much reduced in weight compared with the case of noninteracting carriers. That is, the tendency of a puddle in L1 to electrostatically induce a puddle of opposite charge in L2 is overcompensated by the charge disorder, which exhibits the tendency to induce a puddle of the same charge\footnote{In the (unphysical) limit of vanishing layer separation these tendencies become certainties, but a finite layer separation in concert with the disordered potential landscape allows for local variations of correlation, thereby affecting global correlation measures quantitatively.}. Following the same line of reasoning, we expect maximal anticorrelation if ${R=\infty}$, with the transition from correlation to anticorrelation occuring at some value ${R>1}$. In the following section we substantiate these claims with quantitative predictions for graphene and bilayer graphene.

%\FloatBarrier

\section{\label{Results}Density correlations in double-layers of mono- and bilayer graphene}

We quantify the inter-layer correlations of electron-hole-puddles of the double-layer system described in Sec.~\ref{Bilayersetup} by solving Eqs.~(\ref{nasdef}) and (\ref{Vdef}) self-consistently\footnote{During the self-consistent loop the chemical potential $\mu$ is adjusted such that the mean carrier densities ${\mw{n_j}}$ are kept at a fixed value (zero, unless stated otherwise).} and by comparing the converged carrier densities ${n_1=n(\mathrm{L1})}$ and ${n_2=n(\mathrm{L2})}$ of layers L1 and L2 locally. To that end we calculate the two correlation measures $\xi[n_1,n_2]$ and $\xi_c[n_1,n_2]$, which yield a value of one for perfectly correlated electron-hole puddles (i.e., if the density distribution $n_2$ is proportional to $n_1$ and their values have the same sign at each position $\vec r$), minus one for perfect anticorrelation (i.e., if $n_2$ is proportional to $-n_1$), and are designed for tracking the transition between these two extremes; see Appendix~\ref{AppendixCorrelation} for details.

Figure~\ref{ST3+X_DoubleLayer} depicts potentials and densities for graphene, viz.~linear dispersion, calculated for mean carrier densities ${\mw{n_j}=0}$ and equal strengths of strain and charge disorder (ratio ${R=1}$). Due to the screening effects of the Coulomb interaction within L1, the effective potential for L1 exhibits less variability then the total external potential, i.e., the sum of the disorder potentials $V_{\mathrm{C}}$ and $V_{\mathrm{S}}$. The quasiparticle density $n_1$ of L1 , which can be viewed as resulting from this effective potential, induces an external electrostatic potential for the carrier density $n_2$ of L2. However, for ${R=1}$ the charge disorder potential dominates the total external potential for $n_2$ with magnitudes by a factor of more than 50 larger than those of the electrostatic potential caused by $n_1$. For the setting that leads to Fig.~\ref{ST3+X_DoubleLayer}, the magnitudes of the total external potential for L2 are smaller than those of L1 by a factor of 3--5.
\begin{figure}[H]
\begin{center}
\includegraphics[width=0.85\linewidth]{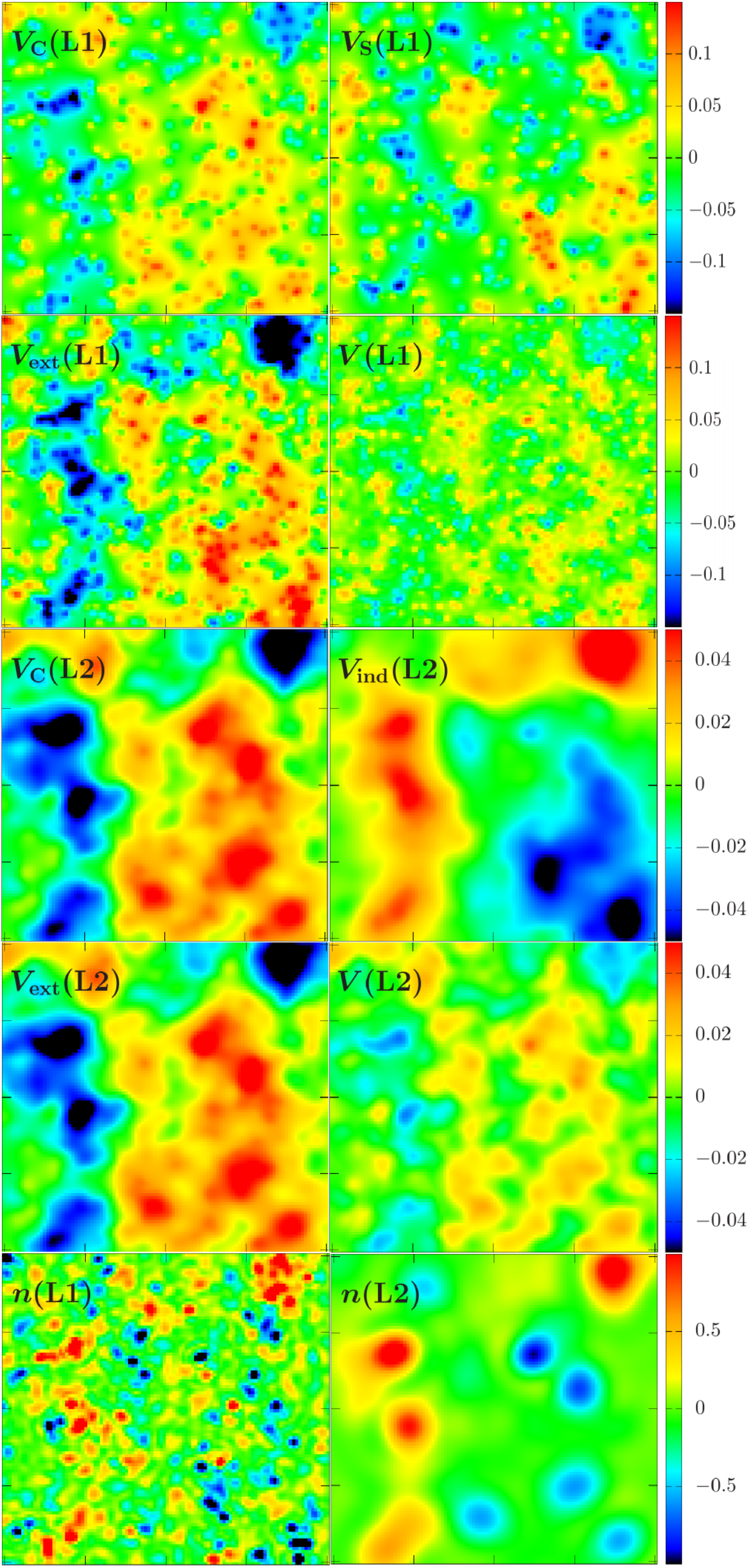}
\caption{\label{ST3+X_DoubleLayer}Visibly correlated carrier densities $n_3$ (including exchange) for layers L1 and L2 (bottom row) from impurity and strain potentials of equal strength ${R=1}$ (top row): This figure illustrates the intermediate potentials relevant in the work flow towards carrier distributions for a double-monolayer graphene system as depicted in Fig.~\ref{DoubleLayerSchematics}. The color codes are in units of $u$ for energy and $l^{-2}$ for density, respectively, and apply to both graphics in each row. Top row: Charged impurity potential $V_{\mathrm{C}}$ (left) and strain potential $V_{\mathrm{S}}$ (right) for L1 --- Second row: Total external potential $V_{\mathrm{ext}}$ (left) and converged effective potential $V$ (right) for L1 ---  Third row:  Charged impurity potential $V_{\mathrm{C}}$ for L2 (left) and electrostatically induced potential $V_{\mathrm{ind}}$ on L2 from the charge distribution $n_1$ of L1 (right, scaled up by a factor of 50) --- Fourth row: Total external potential $V_{\mathrm{ext}}$ (left) and converged effective potential $V$ (right) for L2 --- Bottom row: Converged carrier densities $n_1$ (left, scaled by a factor of 1000) and $n_2$ (right, scaled by a factor of 5000).}
\end{center}
\end{figure}
\noindent As a result, the density fluctuations of $n_2$ are diminished compared with those of $n_1$ by a factor of 5--10. It is therefore well justified to refrain from a self-consistent treatment of the electrostatically induced potentials $V_{\mathrm{ind}}$ of both layers, and to consider only $V_{\mathrm{ind}}$ stemming from $n_1$. As is evident from the bottom row of Fig.~\ref{ST3+X_DoubleLayer}, the spatial distributions of $n_1$ and $n_2$ are correlated rather than anticorrelated for ${R=1}$. 

\begin{figure}[h!]
\begin{center}
\includegraphics[width=0.9\linewidth]{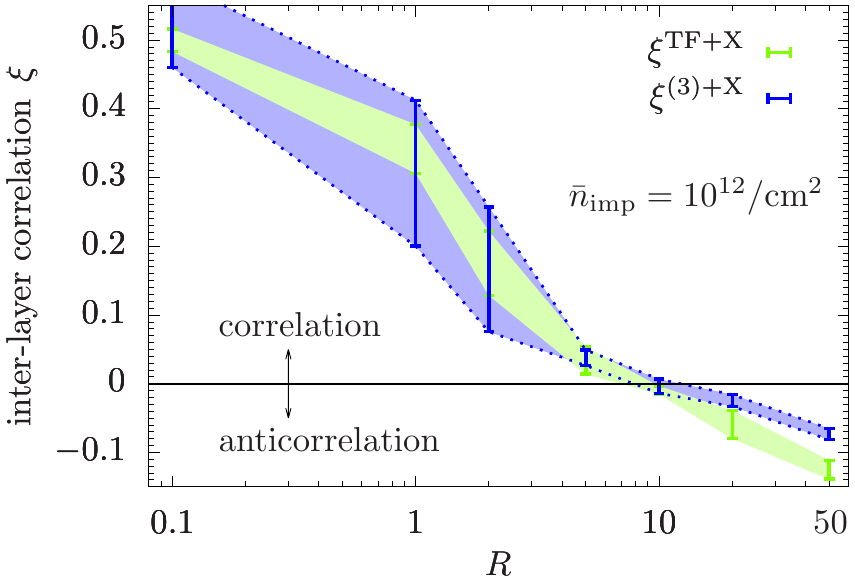}
\caption{\label{Monolayer_graphene_Strain_vs_Charge_MinMax_TF+X_vs_ST3+X_ErrorBars} Both the TF- and quantum-corrected approach predict a transition from correlated to anticorrelated electron-hole-puddles at ${R=R_0^{(3)+\mathrm{X}}\approx R_0^{\mathrm{TF+X}}\approx 10\pm2}$, where $R$ is the ratio of the strength of strain to charged impurity disorder. The correlation measure $\xi$ is plotted as a function of $R$ for five disorder realizations of double monolayer graphene, extracted from both TF- and quantum-corrected carrier densities with exchange incorporated. The average density of charged impurities is ${\bar{n}_{\mathrm{imp}}=10^{12}/\mathrm{cm}^2}$. The error bars demarcate minimal and maximal values of $\xi$ found within the set of disorder realizations; the shaded areas and the dotted line segments guide the eye; the horizontal line at ${\xi=0}$ separates the correlation phase (above) from the phase of anticorrelation (below). We use a logarithmic horizontal axis in order to more clearly showcase the transition point and the contrast of $\xi$ over a large range of $R$.} 
\end{center}
\end{figure}

Repeating the calculation which yields the results illustrated in Fig.~\ref{ST3+X_DoubleLayer} for different values of $R$, we find the critical value $R_0$ at the transition from inter-layer correlation to anticorrelation. The correlation measure $\xi$ signifies inter-layer puddle-correlation (anticorrelation) by taking on positive (negative) values. The corresponding diagram in Fig.~\ref{Monolayer_graphene_Strain_vs_Charge_MinMax_TF+X_vs_ST3+X_ErrorBars}, calculated with charged impurities density $\bar{n}_{\mathrm{imp}}=10^{12}\,\mathrm{cm}^{-2}$, exhibits ${R_0\approx 10\pm2}$ when $n_3$ is used, and a similar value in the case of $n_{\mathrm{TF}}$. Here, we take into account exchange effects and report a rough error estimate simply based on our numerical findings from five disorder realizations\footnote{At this point we do not calculate disorder averages, but rather showcase the predictions of a few disorder realizations (also bearing in mind the limited number of samples usually investigated in actual experiments).}. Evidently, the disparity in electron-hole puddle landscapes between $n_3$ and $n_{\mathrm{TF}}$ as seen in Fig.~\ref{TFvsST3+X_monolayer} does not translate into an appreciable difference between $\xi^{(3)+\mathrm{X}}$ (for $n_3$) and $\xi^{\mathrm{TF+X}}$ (for $n_{\mathrm{TF}}$).

\begin{figure}[h!]
\begin{center}
\includegraphics[width=0.9\linewidth]{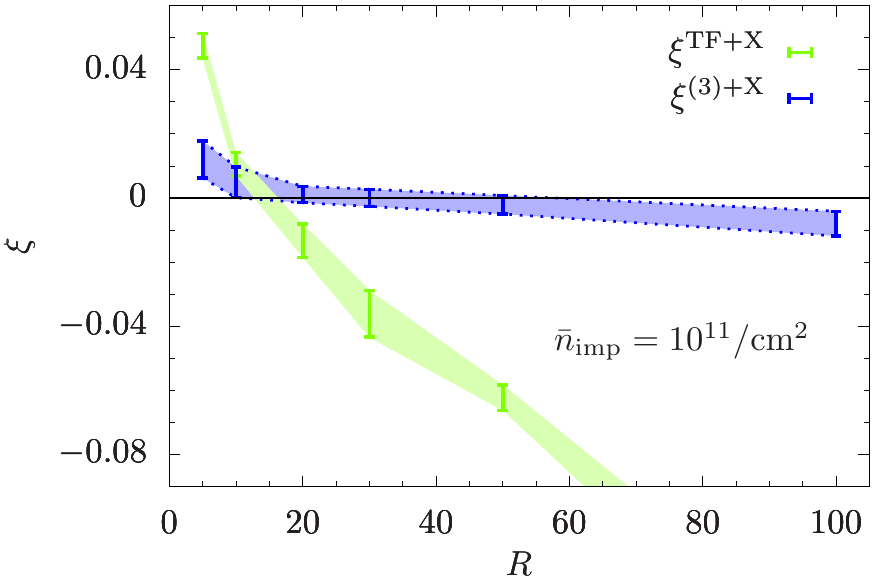}
\caption{\label{Monolayer_graphene_Strain_vs_Charge_MinMax_TF+X_vs_ST3+X_LowNimp2_ErrorBars} At lower values of ${\bar{n}}_{\mathrm{imp}}$, quantum corrections lead to stark differences in the crossover behavior. Here we repeat the plot in Fig.~\ref{Monolayer_graphene_Strain_vs_Charge_MinMax_TF+X_vs_ST3+X_ErrorBars}, but for ${\bar{n}_{\mathrm{imp}}=10^{11}/\mathrm{cm}^2}$, and an increased sheet size to ensure enough disorder statistics (640 impurities on $(800\,\mathrm{nm})^2$). The extended crossover of $\xi^{(3)+\mathrm{X}}$ from correlation to anticorrelation contrasts with the pattern observed in Fig.~\ref{Monolayer_graphene_Strain_vs_Charge_MinMax_TF+X_vs_ST3+X_ErrorBars} and heralds the emergence of quantum--effects (i.e.,  major deviations from $\xi^{\mathrm{TF+X}}$) with decreasing particle numbers.}
\end{center}
\end{figure}

Although it is not surprising per se that integrated quantities like $\xi$ are less sensitive to local differences between $n_3$ and $n_{\mathrm{TF}}$, it cannot be assumed a priori. Our quantitative analysis shows that a quasi-classical approach to inter-layer correlations in monolayer-graphene heterostructures is justified in case of rather large impurity densities like of $10^{12}\,\mathrm{cm}^{-2}$ used for Fig.~\ref{Monolayer_graphene_Strain_vs_Charge_MinMax_TF+X_vs_ST3+X_ErrorBars}. Commonly, however, the TF approximation is less reliable for smaller particle numbers, and quantum corrections can be expected to play a more dominant role as the carrier density is reduced. The disorder potential for smaller $\bar{n}_{\mathrm{imp}}$ is less pronounced and gives rise to puddles that exhibit smaller carrier densities on average. Indeed, with ${\bar{n}_{\mathrm{imp}}=10^{11}/\mathrm{cm}^2}$ employed for Fig.~\ref{Monolayer_graphene_Strain_vs_Charge_MinMax_TF+X_vs_ST3+X_LowNimp2_ErrorBars}, the quantum-corrected $\xi^{(3)+\mathrm{X}}$ can be clearly distinguished from the quasi-classical $\xi^{\mathrm{TF+X}}$. Our data shown in Fig.~\ref{Monolayer_graphene_Strain_vs_Charge_MinMax_TF+X_vs_ST3+X_LowNimp2_ErrorBars} point to the critical value ${R_0^{\mathrm{TF+X}}\approx 15\pm 2}$ with an error estimate similar to that in Fig.~\ref{Monolayer_graphene_Strain_vs_Charge_MinMax_TF+X_vs_ST3+X_ErrorBars}. An increased $R_0$ at lower $\bar{n}_{\mathrm{imp}}$ can be understood from the following simplified picture. With typical values $\left.v_\mathrm{C}\right|_d$ and $v_\mathrm{S}$ of the charge and strain potentials for L1 and neglecting interactions, we have typical values ${n_{\mathrm{TF}}(\mathrm{L1})\propto\big(\mu-v_{\mathrm{ext}}(\mathrm{L1})\big)^2\approx(\left.v_\mathrm{C}\right|_d+v_\mathrm{S})^2}$ for ${\mu\approx0}$, i.e., approximately ${n_{\mathrm{TF}}(\mathrm{L1})\propto v_\mathrm{S}^2}$ at ${R\approx10}$. Then, the typical TF density in L2 is determined by ${v_{\mathrm{ext}}(\mathrm{L2})=\left.v_\mathrm{C}\right|_{\tilde d}+v_\mathrm{ind}}$, where the typical values ${\left.v_\mathrm{C}\right|_{\tilde d}\approx\frac12\left.v_\mathrm{C}\right|_d}$ at ${\tilde{d}-d\approx5}\,$nm and ${v_\mathrm{ind}\propto v_\mathrm{S}^2}$ are roughly equal if ${\bar{n}_{\mathrm{imp}}=10^{12}/\mathrm{cm}^2}$ and ${R=10}$. For a scaled ${\bar{n}_{\mathrm{imp}}=\lambda\times10^{12}/\mathrm{cm}^2}$ and the same ${R=10}$, $v_{\mathrm{ext}}(\mathrm{L1})$ scales with $\lambda$ as well, but the typical values of ${n_{\mathrm{TF}}(\mathrm{L1})}$ then scale like $\lambda^2\,v_\mathrm{S}^2$, such that ${v_{\mathrm{ext}}(\mathrm{L2})=\lambda\left.v_\mathrm{C}\right|_{\tilde d}+\lambda^2\,v_\mathrm{ind}}$. For the case of ${\lambda=1/10}$, as represented by Fig.~\ref{Monolayer_graphene_Strain_vs_Charge_MinMax_TF+X_vs_ST3+X_LowNimp2_ErrorBars}, the strength of $v_\mathrm{ind}$, relative to $\left.v_\mathrm{C}\right|_{\tilde d}$, is diminished by a factor of ten. As strain feeds into $v_\mathrm{ind}$, not into $\left.v_\mathrm{C}\right|_{\tilde d}$, (relatively) more strain is required to counteract the effect of $\left.v_\mathrm{C}\right|_{\tilde d}$, implying a larger critical $R_0$ at lower impurity densities. Disregarding the uncertainties for the critical $R$, one could estimate ${R_0^{(3)+\mathrm{X}}\approx 30}$ for the quantum-corrected correlations. However, $\xi^{(3)+\mathrm{X}}$ rather exhibits an extended cross-over regime, where the magnitude of strain can be varied substantially with little effect on the average correlation $\xi^{(3)+\mathrm{X}}$. Our main result is thus that corrections to the TF approximation can become important even for integrated or averaged observables of 2D materials --- given the proper conditions, for example, small carrier densities.

%\FloatBarrier

%\FloatBarrier

We briefly turn to the impact of exchange effects in inter-layer correlations. Since quantum corrections modify the TF carrier distributions of the individual layers more profoundly than exchange does, cf.~Fig.~\ref{TFvsST3+X_monolayer}, we expect that exchange plays a minor role in determining carrier correlations of double-layer systems. This is confirmed with Table~\ref{Monolayer_graphene_Strain_vs_Charge_R_diagram_withInset_TF_xi_Xnegligible} in Appendix~\ref{AppendixExchange}, where $\xi$ is determined in TF approximation for a single disorder realization, with and without exchange. In view of the magnitude of uncertainties displayed in Figs.~\ref{Monolayer_graphene_Strain_vs_Charge_MinMax_TF+X_vs_ST3+X_ErrorBars} and \ref{Monolayer_graphene_Strain_vs_Charge_MinMax_TF+X_vs_ST3+X_LowNimp2_ErrorBars} we find negligible quantitative differences when omitting or including exchange.

\begin{figure}[htb!]
\begin{center}
\includegraphics[width=0.9\linewidth]{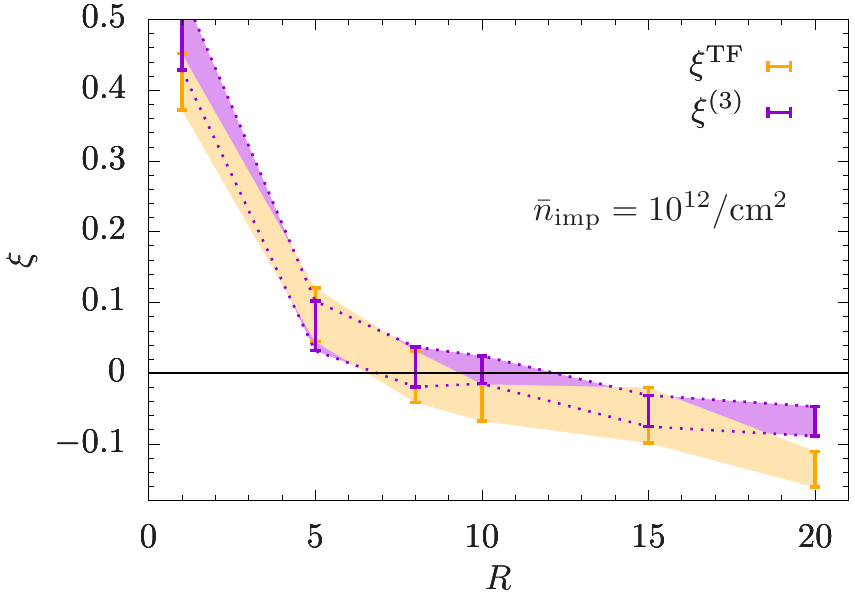}
\caption{\label{Bilayer_graphene_Strain_vs_Charge_MinMax_TF_vs_ST3_ErrorBars} Like Fig.~\ref{Monolayer_graphene_Strain_vs_Charge_MinMax_TF+X_vs_ST3+X_ErrorBars}, but for double bilayer graphene. Using $n_3$, we extract ${R_0^{(3)}\approx 10\pm3}$ for the transition from correlated to anticorrelated carrier densities, but a somewhat smaller value ${R_0^{\mathrm{TF}}\approx 8\pm1}$ in case of the TF approximation. Similar to our observations for double-monolayer graphene, we find that quantum corrections have little effect in the case of double-bilayer graphene for $\bar{n}_{\mathrm{imp}}=10^{12}\,\mathrm{cm}^{-2}$.%\blue{For ${R=\infty}$ we set $V_{\mathrm{C}}=0$ and $R\,W=0.86\,ul$ in Eq.~(\ref{VS}).}
}
\end{center}
\end{figure}

\begin{figure}[htb!]
\begin{center}
\includegraphics[width=0.9\linewidth]{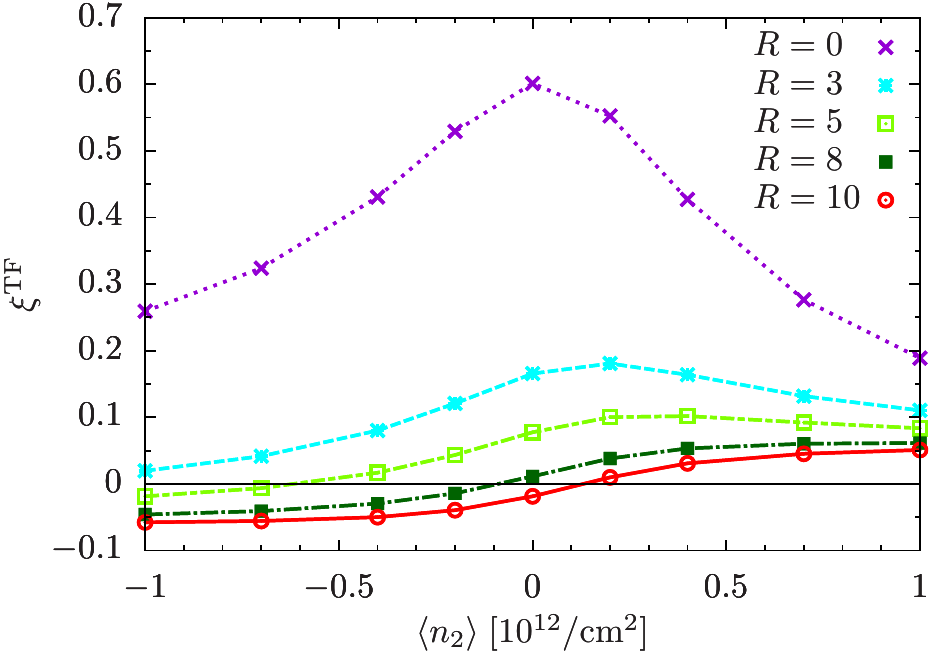}
\caption{\label{xi_n2_R_Bilayer_graphene_TF_200x200_Disorder02} Determination of the critical ratio ${R_0^{\mathrm{TF}}\approx 8}$ for double bilayer graphene (found in Fig.~\ref{Bilayer_graphene_Strain_vs_Charge_MinMax_TF_vs_ST3_ErrorBars}) as a function of $\mw{n_2}$, viz.~$\mu(\mathrm{L2})$. The findings for $\xi^{\mathrm{TF}}$ in Fig.~\ref{Bilayer_graphene_Strain_vs_Charge_MinMax_TF_vs_ST3_ErrorBars} correspond to ${\mw{n_2}=0}$ here. All data points are obtained for a single fixed disorder realization.% Bottom: Schematic depiction of the effective potentials with ${R=0}$ (bottom left, for both L1 and L2) and ${R=8}$ (bottom right, for L2). The chemical potential $\mu$ demarcates the two types of quasiparticles [red(-) and blue(+)] along the dashed horizontal lines, while the horizontal direction indicates spatial extension. In this simplified picture we can evaluate the correlation measure by hand and affirm the phase transition observed in the top figure.
}
\end{center}
\end{figure}

Our results on double bilayer graphene in Fig.~\ref{Bilayer_graphene_Strain_vs_Charge_MinMax_TF_vs_ST3_ErrorBars} show a slight quantitative difference between the transitions of $\xi^{(3)}$ and $\xi^{\mathrm{TF}}$ for the here employed density of charged impurities ${\bar{n}_{\mathrm{imp}}=10^{12}\,\mathrm{cm}^{-2}}$. However, the overall trend for both approximations is still very similar and the spread of $\xi$, originating in multiple disorder realizations like in Figs.~\ref{Monolayer_graphene_Strain_vs_Charge_MinMax_TF+X_vs_ST3+X_ErrorBars} and \ref{Monolayer_graphene_Strain_vs_Charge_MinMax_TF+X_vs_ST3+X_LowNimp2_ErrorBars}, still implies ${R_0^{(3)}\approx R_0^{\mathrm{TF}}}$ within the depicted uncertainties. Extrapolating from the double-monolayer case the double-bilayer graphene can be expected to exhibit more pronounced differences between $\xi^{(3)}$ and $\xi^{\mathrm{TF}}$ for smaller density of charged impurities.

Figure~\ref{xi_n2_R_Bilayer_graphene_TF_200x200_Disorder02} provides another angle on the transition from correlation to anticorrelation for double bilayer graphene: For fixed ${\mw{n_1}=0}$ we chart $\xi^{\mathrm{TF}}$ as a function of $\mw{n_2}$ and find a qualitative change in the shape of the curves $\xi(\mw{n_2})$ as $R$ transits across ${R_0^{\mathrm{TF}}\approx 8}$, consistent with our findings in Fig.~\ref{Bilayer_graphene_Strain_vs_Charge_MinMax_TF_vs_ST3_ErrorBars}.

\FloatBarrier

\section{Conclusions and perspectives}

The main result of this work is thus that both exchange and quantum corrections beyond the Thomas--Fermi approximation are essential for quantitatively reliable density distributions in the context of 2D materials. While exchange plays a minor role for some integrated quantities like inter-layer correlations of double-layer systems, quantum corrections become important for smaller average carrier-densities, viz. cleaner systems. In other words, systems like graphene on boron nitride samples with ${\bar{n}_{\mathrm{imp}} \approx 10^{10} \mathrm{cm}^{-2} }$ call for a more sophisticated description than the Thomas--Fermi approximation can provide. We showed that the regime of high carrier densities may still be described quasiclassically in the computationally highly efficient Thomas--Fermi approximation. For specific systems, however, quantum corrections have to be taken into account at the quantitative level. In this work we provided one such example when analyzing the crossover from correlation to anticorrelation in bilayer heterostructures of 2D materials. We focussed on describing double-layers of mono- and bilayer graphene via recently developed techniques of calculating quantum corrections, but our approach can be easily adapted for other heterostructures.

Turning to the issue of Coulomb drag~\cite{gorbachev_strong_2012}, our analysis reveals that the strain potential has to be at least ten times stronger than the impurity potential in order for anti-correlation between the layers to occur in both monolayer and bilayer graphene. Any strain potential weaker than this will result in correlation. The DPFT framework presented here thus provides the following path towards an experimental verification of the nature of correlations in graphene heterostructures, a non-trivial task since it is impossible to directly measure the local densities of encapsulated 2D layers via local scanning tunneling microscopy. The strength of the strain potential~\cite{gibertini_electron-hole_2012} may first be extracted from local height fluctuation data obtained via atomic force microscopy~\cite{dean_boron_2010}, followed by the impurity strength from conductivity measurements~\cite{adam_self-consistent_2007}. The resulting ratio $R$ may then be compared against Figs.~\ref{Monolayer_graphene_Strain_vs_Charge_MinMax_TF+X_vs_ST3+X_ErrorBars} to \ref{Bilayer_graphene_Strain_vs_Charge_MinMax_TF_vs_ST3_ErrorBars} to determine the nature of correlation between the layers. This would be especially useful for clarifying the source of the unexpected sign changes~\cite{li_negative_2016,lee_giant_2016,simonet_anomalous_2017} in Coulomb drag experiments that have been explained by earlier works simply asserting correlation~\cite{song_energy-driven_2012,schutt_coulomb_2013} or anti-correlation~\cite{gorbachev_strong_2012,ho_theory_2018} without proof. In the case of bilayer graphene, Ref.~[\onlinecite{zarenia_multiband_2018}] theoretically demonstrated that a multiband mechanism based on the thermal smearing of Fermi energy in both layers explains the unexpected sign changes at high densities where puddles may be ignored. A complete analysis at the double neutrality where both puddles and multiband effects are important is however still lacking. In future, this may be addressed by generalizing the DPFT framework presented here to incorporate multiband effects~\cite{zarenia_multiband_2018} in the determination of puddle-induced quantities such as the interlayer correlation and density fluctuations.

Finally, the findings of this work are useful in the study of van der Waals heterostructures~\cite{geim_van_2013} consisting of a pair of two-dimensional electronic layers separated by thin dielectric spacers.
These structures serve as ideal platforms for the study of a range of interesting physical effects such as exciton condensation~\cite{li_excitonic_2017,liu_quantum_2017,burg_strongly_2018,lopez_rios_evidence_2018} and strong light-matter interaction~\cite{britnell_strong_2013,yu_highly_2013,withers_light-emitting_2015} and are thus an area of intense research activity. 
It is likely that the inter-layer puddle correlations calculated within this work plays a role in the above physics.

\acknowledgments
We acknowledge the support of the National Research Foundation of Singapore under its Fellowship program (NRF-NRFF2012-01) and the National University of Singapore Young Investigator Award (R-607-000-094-133). We also acknowledge the use of the dedicated computational facilities at the Centre for Advanced 2D Materials.

\appendix

\section{\label{AppendixUnits} Units and system parameters}

Throughout this work we use cgs units, measure lengths in ${l=\sqrt{3}a\approx 2.46\,\mbox{\r{A}}}$, where $a$ is the lattice constant of graphene, energies in ${u=\hbar\vF/l\approx2.678\,\mathrm{eV}}$ (for monolayer graphene, with the Fermi velocity {${\vF=10^6}\,$m/s}) and {${u'=\hbar^2/(m\,l^2)\approx 36.71}\,$eV} (for bilayer graphene with quasiparticle mass~\cite{Raza2012} {${m\approx 0.39\,\mathrm{eV}/(2\,\vF^2)}$}). With the elementary charges in cgs- and SI-units connected via ${\mathrm{e}_{\mathrm{SI}}=\sqrt{4\pi}\mathrm{e}_{\mathrm{cgs}}}$ and the dielectric constants related via ${\epsilon=\kappa\epsilon_0}$, the Coulomb coupling strength in Eqs.~(\ref{HartreeEnergy}) and (\ref{ExchangePotentialRossi}) reads
\begin{align}
W=\frac{\mathrm{e}_{\mathrm{cgs}}^2}{\kappa\epsilon_0}=\hbar\vF r_\mathrm{s}\approx\frac{2.187}{\kappa}ul
\end{align}
for graphene and
\begin{align}
W\approx\frac{0.1596}{\kappa}u'l
\end{align}
for bilayer graphene. The static limit~\cite{Geick1966} for h-BN amounts to setting ${\kappa=5.09}$.

We restrict the numerical evaluations to square-shaped sheets of 2D materials with edge length of ${L=200\,\mathrm{nm}=406.5\,l}$, unless stated otherwise. A graphene sheet of $(200\,\mathrm{nm})^2$ consists of approximately 1.5 million carbon atoms, while the number of charge carriers, derived from the $n_3$-type realizations of $n_\pm(\vec r)$, typically ranges between 100 and 10000 for layer L1 --- depending on the employed disorder strengths. We confirmed the convergence of the correlation measure $\xi$ by increasing the resolution --- here a modest grid size of $105^2$ grid points proved enough for converging $\xi$ sufficiently.

The main contribution to the quantum-corrected density $n_3(\vec r)$ stems from the effective potential $V$ in the vicinity of $\vec r$, see Table~\ref{TableDensities}, but the spatial integral for $n_3$ is a priori indefinite. We restrict this spatial integral to the square-shaped sheet of area $L^2$. Hence, the bulk of the sheet is modeled adequately, but the region close to the edges requires further consideration. Two scenarios suggest themselves. First, the Coulomb tails of the charged impurities could be taken into account beyond the here employed sheet area $L^2$ (which would increase the computational cost substantially, with little effect on $\xi$). For this case of a truly finite sheet also the linear (quadratic) dispersion for graphene (bilayer graphene) would require a modification due to the non-periodic boundary conditions. Second, we could view the sheet as a representative part of a larger sample whose disorder potential is unknown beyond the employed sheet area\footnote{An extension of the position space support by employing appropriate reflections of the disordered sheet at its edges, would counteract this inaccuracy, but likely introduce spurious effects due to the artificial periodicity, along with a substantial increase in computational cost.}. That is, both scenarios render the $n_3$ densities less precise near the edges (and particularly near the corners) of the sheet when omitting the spatial regions beyond the employed area $L^2$ (which amounts to setting ${V(\vec z)=\mu}$ for ${\vec z\notin L^2}$). We deem this inaccuracy acceptable  since the majority of the sheet is adequately taken into account, and the effect on $\xi$ is likely irrelevant given the observed variation due to different disorder realizations.

\section{\label{AppendixExchange} Exchange}

\begin{figure}[htb!]
\begin{center}
\includegraphics[width=0.95\linewidth]{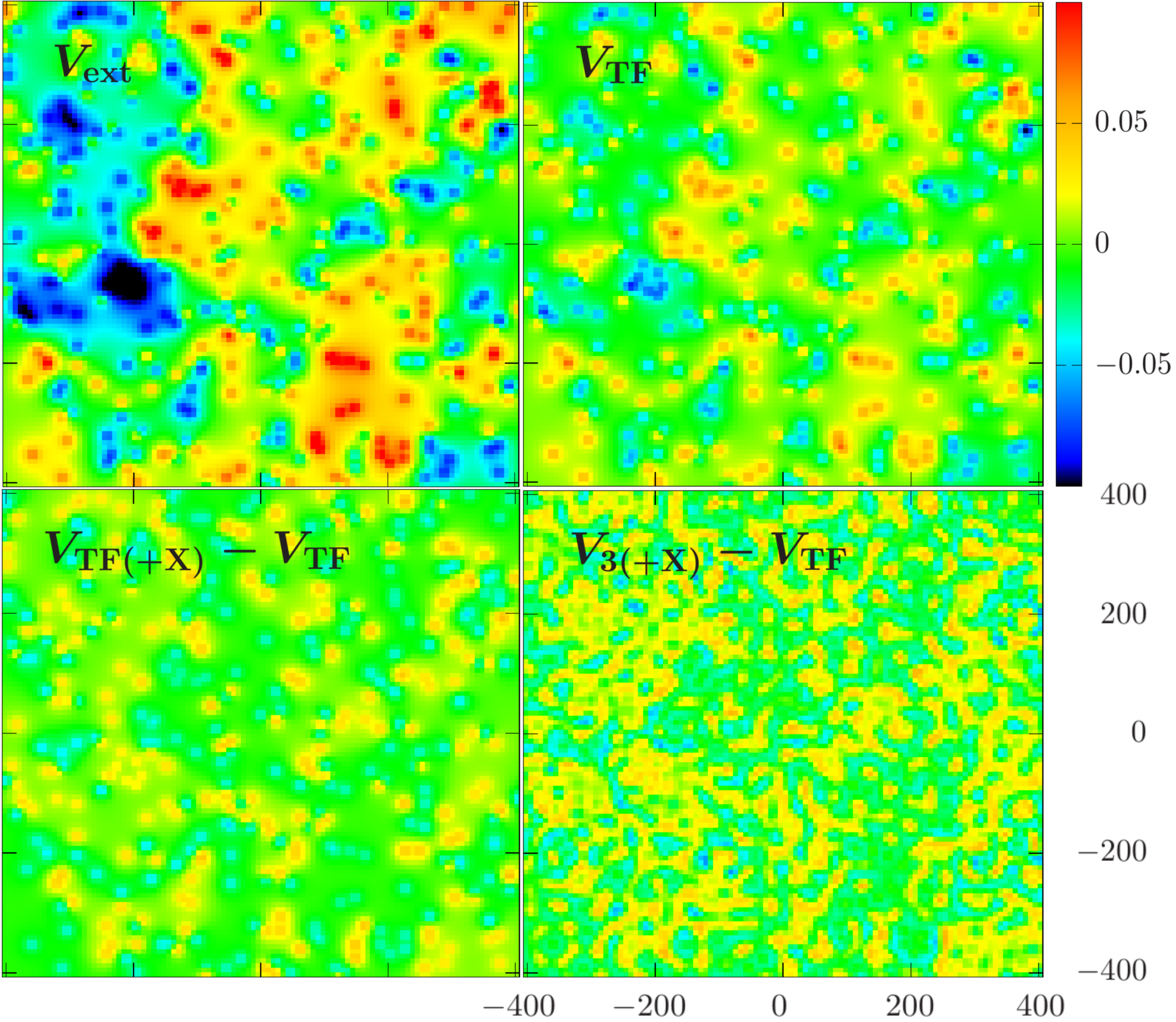}
\caption{\label{TFpotentials_monolayer}Quantum corrections to the TF approximation are visibly significant also at the level of effective potentials. Here, we depict the external potential $V_{\mathrm{ext}}$ for a graphene sheet of size $L^2$, corresponding to one realization of charged impurity disorder (top left), the self-consistently converged effective potential $V$ in TF approximation (denoted $V_{\mathrm{TF}}$ here) with the Hartree term only (top right), the differences between $V_{\mathrm{TF}}$ and $V_{\mathrm{TF}(+X)}$, where the Hartree term plus the exchange of Eq.~(\ref{ExchangePotentialRossi}) consitutes the interaction potential (bottom left), and the comparison of the quantum-corrected effective potential (including exchange) with $V_{\mathrm{TF}}$ (bottom right). The color code presents energy in units of $u$. The bottom panels are scaled by a factor of $5$.}
\end{center}
\end{figure}

Exchange potentials for graphene are available, for example in Ref.~[\onlinecite{Rossi2008}],
\begin{align}\label{ExchangePotentialRossi}
V_{\mathrm{X}}(\vec r)&\approx W\sqrt{\pi|n(\vec r)|}\,\mathrm{sgn}\big(n(\vec r)\big)\nn\\
&\quad\times\left[\frac14\log\left(\frac{4\,\sqrt{3}}{\sqrt{\pi|n(\vec r)|}}\right)-0.5757\right]
\end{align}
in the units used here. Figure~\ref{TFpotentials_monolayer} illustrates the exchange effects on the converged effective potential $V$, and Table~\ref{Monolayer_graphene_Strain_vs_Charge_R_diagram_withInset_TF_xi_Xnegligible} demonstrates that exchange is negligible as far as inter-layer correlations are concerned.

An approximate exchange potential for other materials, viz.~ other dispersion relations, can for instance be obtained by calculating the according Dirac exchange energy for 2D systems and adding correlation in the spirit of Ref.~[\onlinecite{Vosko1980}]. It would also be interesting to see if exchange-correlation functionals developed for the 2D electron gas like in Ref.~[\onlinecite{Pittalis2010}] can be used for systems with quadratic dispersion differing in the particle mass, e.g.~for effective bilayer graphene. This topic is, however, beyond the scope of this work.

\begin{table}
\caption{\label{Monolayer_graphene_Strain_vs_Charge_R_diagram_withInset_TF_xi_Xnegligible} The correlation measure $\xi$, given in Eq.~(\ref{CMxi}), as a function of the ratio $R$ between strain and disorder strength (cf.~Eq.~(\ref{VS})) for one disorder realization of double monolayer graphene in TF approximation, with and without exchange (X). In absolute numbers on the relevant scale of $[-1,1]$ for the correlation measures and regarding the spread of $\xi$ due to different disorder realizations, cf. Fig.~\ref{Monolayer_graphene_Strain_vs_Charge_MinMax_TF+X_vs_ST3+X_ErrorBars}, we find exchange effects to be insignificant for investigating correlations between the two layers L1 and L2. We made the same observation for different disorder realizations.\\[-0.5em]}
\begin{tabular}{c|l|l|l|l|l}
$R$ & 0.1 & 1 & 10 & 20 & 50 \\
\hline\hline\\[-2.0ex]
$\xi^{\mathrm{TF+X}}$ & 0.492 & 0.347 & -0.013 & -0.051 & -0.122 \\
$\xi^{\mathrm{TF}}$ & 0.480 & 0.341 & -0.015 & -0.058 & -0.135
\end{tabular}
\end{table}

\FloatBarrier

\section{\label{AppendixDisorder} Disorder potentials}

We separate the sheet (with optional strain) of the first layer L1 from the plane that holds the charged impurities by ${d=1}\,$nm, cf.~Fig.~\ref{DoubleLayerSchematics}. The impurity-induced Coulomb potential in a sheet at distance $d$ reads
\begin{align}\label{AD01}
V_\mathrm{C}(\vec r)=W\int(\d\vec r')\frac{C(\vec r')}{\sqrt{|\vec r-\vec r'|^2+d^2}},
\end{align}
where $C$ takes on values ${c_i\in\{-1,0,1\}}$ randomly on the grid points ${i\in\{1,\dots,\Lambda\}}$. We employ an average density of charged impurities $n_{\mathrm{imp}}$ (e.g., $10^{12}\,\mathrm{cm}^{-2}$, corresponding to 400 impurities on $L^2$), with half of the impurities positively charged, such that the net charge in the impurity plane is zero. We model the strain as
\begin{align}\label{VS}
V_\mathrm{S}(\vec r)=R\,W\int(\d\vec r')\frac{\tilde{C}(\vec r')}{\sqrt{|\vec r-\vec r'|^2+d^2}},
\end{align}
with $R$ controlling the relative impact of strain and charged impurities, while $\tilde{C}$ and $C$ represent different disorder realizations of the same type. The sum of Eqs.~(\ref{AD01}) and (\ref{VS}) is the external potential $V_\mathrm{ext}$, cf.~Eq.~(\ref{V-}), for a monolayer setup as well as for the first sheet of the double-layer structure. The second sheet of the double-layer structure is not affected by $V_\mathrm{S}$, but only by the charged impurities at a distance of ${\tilde{d}=6}\,$nm, cf.~Fig.~\ref{DoubleLayerSchematics}, and by the electrostatic potential from the converged carrier distribution $n_1$ on the first sheet:
\begin{align}\label{tildeVext}
V_{\mathrm{ext}}(\mathrm{L2})(\vec r) = \left.V_\mathrm{C}(\vec r)\right|_{d=\tilde{d}}+\int(\d\vec r')\frac{W\,n_1(\vec r + \vec r')}{\sqrt{|\vec r'|^2+(\tilde{d}-d)^2}}.
\end{align}

\section{\label{AppendixCorrelation} Correlation measures}

In Section~\ref{Results} we use a normalized version of the interlayer correlation measure~\cite{rodriguez-vega_ground_2014}
\begin{align}\label{CMxi}
\xi[n_1,n_2]=\frac{\mw{(n_1-\mw{n_1})(n_2-\mw{n_2})}}{n_{1,\mathrm{rms}}\,n_{2,\mathrm{rms}}}
\end{align}
to quantify the degree of (anti-)correlation between the density distributions $n_1$ and $n_2$. Here, $\mw{n_j}=\frac{1}{\Lambda}\sum_{i=1}^{\Lambda}n_j(i)$ is the mean of the discretized density of layer $j$ on the grid points ${i\in\{1,\dots,\Lambda\}}$, and the root mean square of the density fluctuations in layer $j$ is ${n_{j,\mathrm{rms}}=\sqrt{\mw{(n_j-\mw{n_j})^2}}}$.
\begin{figure}[H]
\begin{center}
\includegraphics[width=0.9\linewidth]{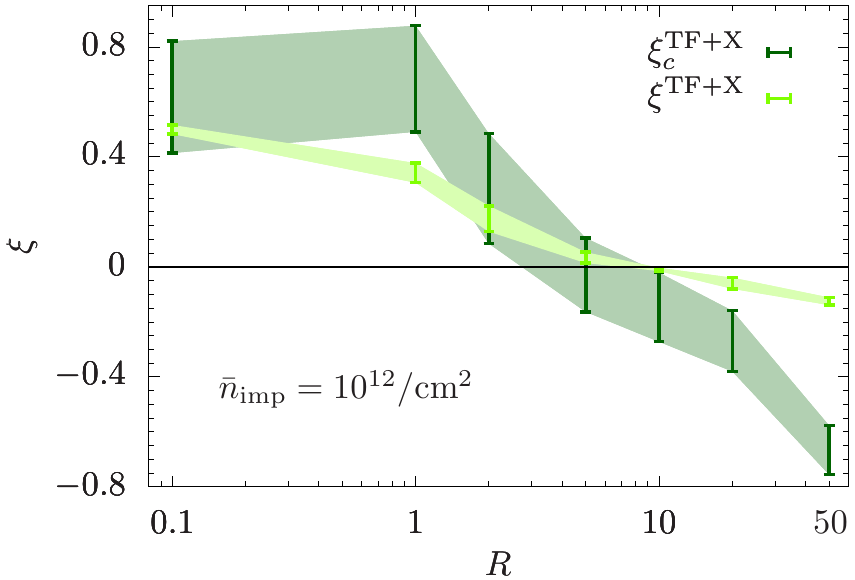}
\caption{\label{Monolayer_graphene_Strain_vs_Charge_MinMax_TF+X_xi_vs_TF+X_xic_ErrorBars}Comparison between the correlation measures $\xi$ and $\xi_c$ for the TF densities, including exchange. As expected, $\xi_c$ provides more contrast than $\xi$, but its increased variance renders it less useful for our purposes.}
\end{center}
\end{figure}
To check for spurious artefacts of the correlation measure $\xi$ in Eq.~(\ref{CMxi}) and provide an error estimate for the (anti-)correlations, we also consider the alternative measure
\begin{align}\label{CMxic}
\xi_c[n_1,n_2]=\frac{\mw{\,(n_1-\mw{n_1}_c)(n_2-\mw{n_2}_c)}_c}{n_{1,\mathrm{rms},c}\,n_{2,\mathrm{rms},c}},
\end{align}
which yields stronger contrast between situations of correlation and anticorrelation --- at the expense of larger variance; cf.~Fig.~\ref{Monolayer_graphene_Strain_vs_Charge_MinMax_TF+X_xi_vs_TF+X_xic_ErrorBars}. Here we define the average
\begin{align}
\mw{A}_c=\frac{\sum_{i=1}^{\Lambda}\chi_c(i)A(i)}{\sum_{i=1}^{\Lambda}\chi_c(i)}
\end{align}
of a quantity $A$ and the corresponding root mean square $n_{j,\mathrm{rms},c}=\sqrt{\mw{(n_j-\mw{n_j}_c)^2}_c}$. The characteristic function $\chi_c(i)$ at grid point $i$ equals one if ${n_j(i)\notin I_j}$ (for both ${j=1,2}$) and zero otherwise\footnote{All grid points are taken into account, i.e., ${\chi_c(i)=1}$ for all $i$, if ${c=1/2}$. Then, Eqs.~(\ref{CMxi}) and (\ref{CMxic}) coincide. For ${c=0}$ all data is dismissed.} --- with the cutoff interval $I_j=[\mathrm{min}_i\{n_j(i)\}+c\,\Delta_j,\mathrm{max}_i\{n_j(i)\}-c\,\Delta_j]$ and the density spread ${\Delta_j=\mathrm{max}_i\{n_j(i)\}-\mathrm{min}_i\{n_j(i)\}}$. We choose ${c=0.4}$ to dismiss the fluctuations of correlation at small local carrier density. Owing to its increased variance, we refrain from depicting $\xi_c$ in Sec.~\ref{Results}.

\section{\label{AppendixSCloop} Self-consistent ground-state variables}

\begin{figure}[h!]
\begin{center}
\includegraphics[width=0.9\linewidth]{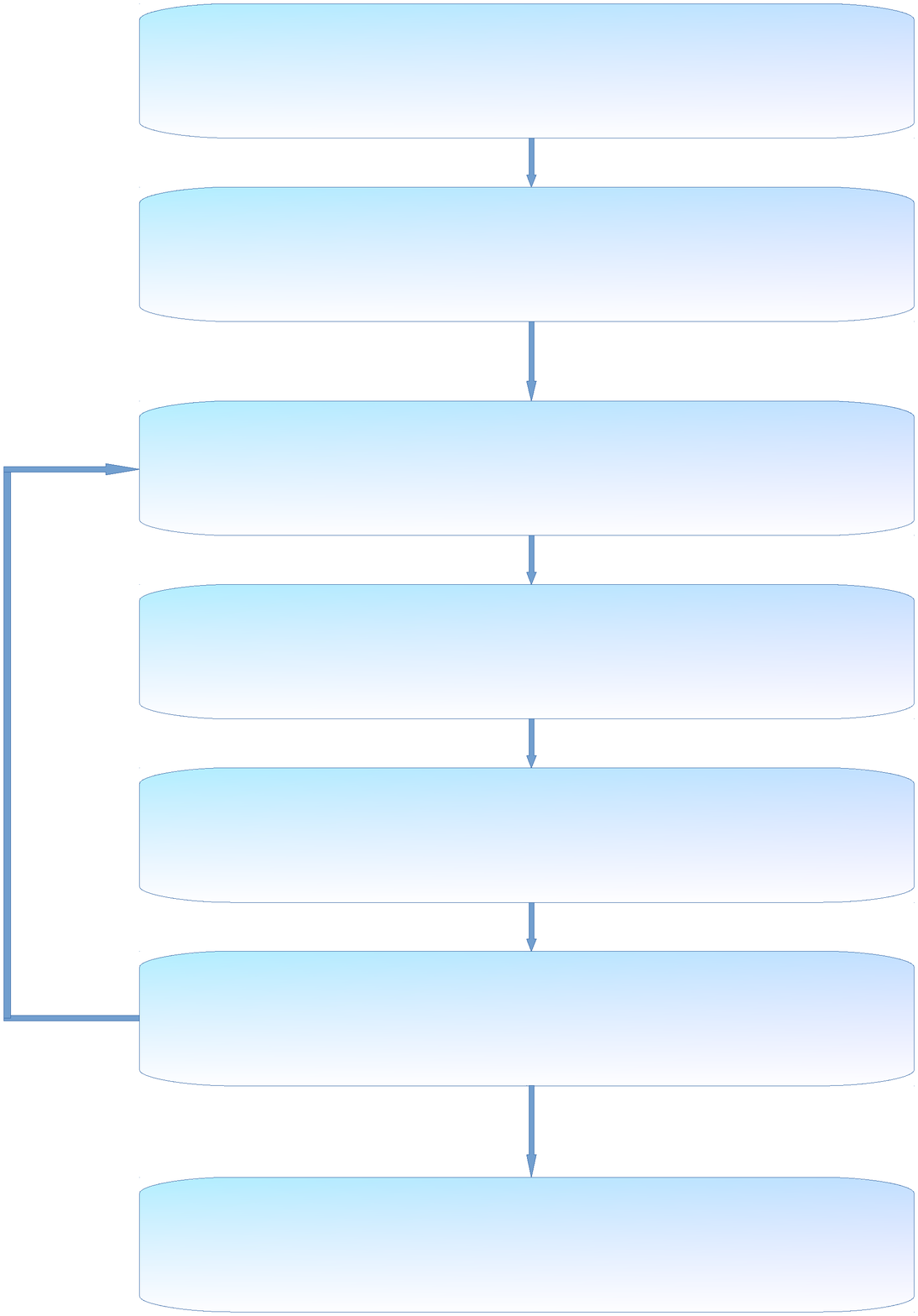}
\put(-178,297){(I) Initialize noninteracting densities $n[V-\mu]$}
\put(-178,287){\phantom{(I) }with any chemical potential $\mu$}
\put(-178,253){\phantom{xx}(II) Adapt $\mu$ to enforce choice of particle}
\put(-178,243){\phantom{xx(II) }number $N$}
\put(-178,203){\phantom{xxxx}(III) Update effective potentials $V[n]$}
\put(-178,193){\phantom{xxxx(III) }with densities}
\put(-178,161){\phantom{xx}(IV) Update densities $n[V]$ with effective}
\put(-178,151){\phantom{xx(IV) }potentials and adapt $\mu$ to enforce $N$}
\put(-178,112){\phantom{xxxxxx}(V) Mix old and new densities}
\put(-178,70){\phantom{xxxxxx}(VI) Densities have converged?}
\put(-178,16){\phantom{xxxxx}Stationary solutions $n$, $V$, $\mu$ found}
\put(-88,43){yes}
\put(-205,75){no}
\caption{\label{FlowChart}Schematic of the work flow for implementing the self-consistent loop of Eqs.~(\ref{nasdef})--(\ref{Nmu}). Steps (I) and (II) represent the initialization. The self-consistent loop is carried out in steps (III)-(VI).}
\end{center}
\end{figure}

In this appendix we outline the numerical procedures for obtaining the ground-state variables of DPFT, viz.~stationary solutions, $n$, $V$, and $\mu$ of Eqs.~(\ref{nasdef})--(\ref{Nmu}), adapted to the quasi-particle system described in Sec.~\ref{SCsim}. Figure~\ref{FlowChart} illustrates the main steps for implementing the self-consistent loop, with a detailed description below.\\[1\baselineskip]

\noindent
\underline{Initialization ${(\mbox{iteration }i=0)}$:}\\
\begin{itemize}
 \item[(I)] Calculate ${n_\pm^{(0)}=n_\pm^{(i=0)}=n_\nu[V_\pm^{(0)}-\mu^{(0)}]}$ using ${V_-^{(0)}=V_{\mathrm{ext}}}$ from Eq.~(\ref{V-}) and $V_+^{(0)}$ from Eq.~(\ref{V+}). The initial choice of the chemical potential is arbitrary, e.g., ${\mu^{(0)}=0}$. We omit $\vec r$-dependences in this appendix for the sake of notational simplicity. ${n_\nu[V-\mu](\vec r)}$ are functionals of ${V-\mu}$ with parametric dependence on $\vec r$, for example, the density expressions given in Table~\ref{TableDensities}.
 \item[(II)] Adapt $\mu^{(0)}$ until ${N=\int(\d\vec r)\,n^{(0)}=\int(\d\vec r)\,(n_-^{(0)}-n_+^{(0)})}$. For instance, we demand ${N=0}$ to be reached within an absolute accuracy of $10^{-4}$.
\end{itemize}
\noindent\underline{Self-consistent loop:}
\begin{itemize}
 \item[(III)] Update effective potential $${V_-^{(i+1)}=V_{\mathrm{ext}}}+\left.\frac{\delta E_{\mathrm{int}}[n]}{\delta n}\right|_{{n_-=n_-^{(i)}}}$$ and determine effective potential ${V_+^{(i+1)}}$ via Eq.~(\ref{V+}).
 \item[(IV)] Update densities ${\tilde n_\pm^{(i+1)}:=n_\nu[V_\pm^{(i+1)}-\mu^{(i+1)}]}$, starting with ${\mu^{(i+1)}=\mu^{(i)}}$, and adjusting $\mu^{(i+1)}$ until ${N=\int(\d\vec r)\,\tilde n^{(i+1)}=\int(\d\vec r)\,(\tilde n_-^{(i+1)}-\tilde n_+^{(i+1)})}$.
 \item[(V)] Mix old and new densities with mixing parameter ${\theta\in(0,1)}$: ${n^{(i+1)}=(1-\theta)\,n^{(i)}+\theta\,\tilde n^{(i+1)}}$. More sophisticated density mixing like Pulay or Broyden mixing can be expected to improve the convergence behaviour.
 \item[(VI)] Check convergence with an appropriate norm criterion ${||n^{(i+1)}-n^{(i)}||<\epsilon}$. We use ${\epsilon=10^{-6}}$ and
 \begin{align}
  ||n^{(i+1)}-n^{(i)}||&=\frac{\sum_{\vec r_j}\chi_{ij}\,|n^{(i+1)}(\vec r_j)-n^{(i)}(\vec r_j)|}{s^{(i+1)}\sum_{\vec r_j}\chi_{ij}},
  \end{align}
 with grid positions $\vec r_j$,
 \begin{align}
  \chi_{ij}&=\eta\left(t^{(i+1)}(\vec r_j)-\frac{s^{(i+1)}}{1000}\right),\\
  t^{(i+1)}(\vec r_j)&=\mathrm{min}\{|n^{(i+1)}(\vec r_j)|,|n^{(i)}(\vec r_j)|\},\\
  s^{(i+1)}&=\frac12\big(\mathrm{MAX}_i-\mathrm{MIN}_i\big),\\
  \mathrm{MAX}_i&=\mathrm{max}_{\vec r}\big\{\mathrm{max}\{|n^{(i+1)}(\vec r)|,|n^{(i)}(\vec r)|\}\big\},\\
  \mathrm{MIN}_i&=\mathrm{min}_{\vec r}\big\{\mathrm{min}\{|n^{(i+1)}(\vec r)|,|n^{(i)}(\vec r)|\}\big\}.
 \end{align}
 If densities have converged to within $\epsilon$, the stationary solutions $n$, $V$, and $\mu$ of Eqs.~(\ref{nasdef})--(\ref{Nmu}) are found; otherwise return to step (III) for the next iteration.
\end{itemize}
The number of self-consistent iterations $i$ required for convergence is largely determined by the density-mixing parameter $\theta$, which is highly system-dependent. Here, we can choose ${\theta=0.5}$ for weak disorder, but need to approach the equilibrium very gently in case of strong disorder ($\theta\lesssim0.01$). In our implementation $i$ usually reaches values on the order of $\sim 10/\theta$.

\section{\label{Appendixn3} The quantum-corrected density $n_3$\\ for linear dispersion}

In this appendix, we outline the derivation of the density expressions for linear dispersion given in Table~\ref{TableDensities}. Equations~(\ref{nasdef}) and (\ref{E1}) yield Eq.~(\ref{nPropagator}) with the aid of the Fourier transform of the step function $\eta(\cdot)$:
\begin{align}
n(\vec r)&=g\bok{\vec r}{\eta(\mu-H_1)}{\vec r}=g\Int\frac{\d t}{2\pi\I t}\,\Exp{\frac{\I t}{\hbar}\mu}\bra{\vec{r}}\Exp{-\frac{\I t}{\hbar}H_1}\ket{\vec{r}},
\end{align}
where the degeneracy factor $g$ is part of the trace in Eq.~(\ref{E1}). Generally, a Suzuki-Trotter factorization $U_\nu$ of the time-evolution operator with
\begin{align}
\bok{\vec r}{U_\nu}{\vec r}&=\bok{\vec r}{\prod_{i=1}^{\nu/2}\mathrm{e}^{\alpha_i T(\vec P)}\mathrm{e}^{\beta_i V(\vec R)}}{\vec r}\\
&=:\prod_{i=1}^{\nu/2}\int(\d\vec r_i)(\d\vec p_i)\, c_i\,\mathrm{e}^{-\frac{\I t}{\hbar}A}
\end{align}
and real-valued functions $c_i$ and $A$ fixed by the here inserted completeness relations in position and momentum space, implies
\begin{align}
\bok{\vec r}{\eta(\mu-H_1)}{\vec r}&=\prod_{i=1}^{\nu/2}\int(\d\vec r_i)(\d\vec p_i) \,c_i\,\eta(\mu-A).
\end{align}
For example, the case $U_3$ in Sec.~\ref{DPFT} is formally obtained by setting ${\alpha_1=\alpha_2=-\frac{\I t}{2\hbar}}$, ${\beta_1=-\frac{\I t}{\hbar}}$, and ${\beta_2=0}$ in $U_4$. Note that $U_3$ is denoted as $U_{3'}$ in Ref.~[\onlinecite{Chau2018}].

We obtain the TF density by choosing ${\nu=2}$ and ${\alpha_1=\beta_1=-\frac{\I t}{\hbar}}$. Specifying linear dispersion, ${T(\vec p)=\vF|\vec p|}$, we find
\begin{align}
n_{\mathrm{TF}}(\vec r)&=2\pi g\int_0^\infty\frac{\d p\,p}{(2\pi\hbar)^2}\eta\big(\mu-\vF\,p-V(\vec r)\big),
\end{align}
whose analytic result is easily obtained and given in Table~\ref{TableDensities}. The computation of $n_3$ requires the insertion of three completeness relations, and the so introduced two-dimensional momentum integrations are reduced to one-dimensional integrals via appropriate coordinate transformations. The subsequent analytic evaluations of the remaining one-dimensional integrals lead to the expression for $n_3$ in Table~\ref{TableDensities}, leaving one spatial integral to be evaluated numerically.

%\FloatBarrier

%\bibliographystyle{ap­srev4-1}

\bibliography{/home/martintrappe/Desktop/PostDoc/PapersTalksPoster/myPostDocbib}
%\bibliography{myPostDocbib}

\end{document}